\documentclass[aps,prb,citeautoscript,longbibliography,superscriptaddress,showpacs,twocolumn]{revtex4-1} 

\usepackage{enumerate}
\usepackage{amssymb}
\usepackage{amsmath,dsfont}
\usepackage[usenames]{color}
\usepackage{braket, pifont}
\usepackage{graphicx,bm}
\usepackage[colorlinks,citecolor=blue]{hyperref}

\allowdisplaybreaks
\begin{document}
\title{Transverse profile and 3D spin canting of a Majorana state in carbon nanotubes}

\author{Lars Milz}
\affiliation{Institute for Theoretical Physics, University of Regensburg, 93053 
Regensburg, Germany} 
\author{Wataru Izumida}
\affiliation{Institute for Theoretical Physics, University of Regensburg, 93053 
Regensburg, Germany} 
\affiliation{Department of Physics, Tohoku University, Sendai 980 8578, 
Japan} 
\author{Milena Grifoni}
\affiliation{Institute for Theoretical Physics, University of Regensburg, 93053 
Regensburg, Germany} 
\author{Magdalena Marganska}
\thanks{Corresponding author. Email: magdalena.marganska@ur.de}
\affiliation{Institute for Theoretical Physics, University of Regensburg, 93053 
Regensburg, Germany} 

\begin{abstract}
The full spatial 3D profile of Majorana bound states (MBS) in a nanowire-like setup featuring a semiconducting carbon nanotube (CNT) as the central element is discussed. By atomic tight-binding calculations we show that the chiral nature of the CNT lattice is imprinted in the MBS wave function which has a helical structure, anisotropic in the transverse direction. The local spin canting angle displays a similar spiral pattern, varying around the CNT circumference. 
We reconstruct the intricate 3D profile of the MBS wave function analytically, using an effective low energy Hamiltonian accounting both for the electronic spin and valley degrees of freedom of the CNT. In our model the four components of the Majorana spinor are related by the three symmetries of our Bogoliubov-de Gennes (BdG) Hamiltonian, reducing the number of independent components to one.  
A Fourier transform analysis uncovers the presence of three contributions to the MBS, one from the $\Gamma$-point and one from each of the Fermi points, with further complexity added by the presence of two valley states in each contribution. 
\end{abstract}
%
% \begin{abstract}
% The Majorana nanowire setup is a versatile one, where both the bulk superconductor and the central quasi-1D system can be chosen according to the researchers' needs. 
% An advantage of replacing the nanowire with a semiconducting carbon nanotube is that we can set up its tight-binding model and solve it numerically.
% The results of this {\em in silico} experiment confirm the possibility of realizing Majorana bound states (MBS) in this setup, and have been reported in our recent paper.
% Here we extend this study by constructing analytically a full 3D spatial profile of the Majorana wave function, matching the numerical results, with the help of an effective two-band model. The MBS turns out to contain three momentum contributions, one from the $\Gamma$ point and one from each of the Fermi points. Moreover, the presence of two valleys results in an intricate helical pattern of the wave function that cannot be factorized into separate transverse and longitudinal profile.
% Knowing the full wave function, we calculate the local spin canting angle of the Majorana bound state, which displays a spiral pattern, varying around the nanotube's circumference. This will affect the electron tunneling into the Majorana state, though whether the overall effect will be favorable or deleterious remains yet to be seen. 
% We also show that all components of the Majorana spinor are related by the three symmetries of the Bogoliubov-de Gennes Hamiltonian, reducing the number of independent components to one.
% \end{abstract}

\maketitle
\noindent
Over the past decade Majorana fermions have been of great interest in condensed matter physics.  Under special conditions they arise as quasiparticles in superconductors,\cite{Aguado-RNC-2017} where they are zero energy eigenstates of the Bogoliubov-de Gennes (BdG) Hamiltonian and of the particle-hole symmetry operator. Theoretically such quasiparticles were predicted to appear in the elusive one-dimensional $p$-wave superconductors \cite{Kitaev-PhysUsp-2001}; but it is also possible to engineer $s$-wave systems in such a way that they mimic $p$-wave superconductivity \cite{Sato-RPP-2017}. The most popular setup is based on semiconducting nanowires with large spin-orbit interaction and large $g$-factor in contact with a superconductor, which induces superconducting proximity correlations in the wire \cite{Lutchyn-PRL-2010,Oreg-PRL-2010}. Although the experiments are by now  very advanced  \cite{Lutchyn-NatRevMat-2018}, a definite proof that the reported signatures~\cite{Mourik-Science-2012,Churchill-PRB-2013,Deng-Science-2016,Zhang-NatComm-2017}  are  really due to the topologically non trivial Majorana bound states  (MBS) is still missing. Thus, recent proposals have suggested to use local probes to infer exclusive properties of a MBS, such as its nonlocality and its peculiar spin canting structure\cite{Liu-PRB-2017,Prada-PRB-2017, Clarke-PRB-2017,Deng-NatPhys-2017,Hoffman-PRB-2017,Schuray-PRB-2018}, or the maximal electron-hole content of the Majorana spinor \cite{Bena-PRL-2012,Bena-PRB-2015}. However, in order to exclude spurious effects, local experiments can be truly useful only if the spatial profile of the MBS is known with sufficient accuracy. This is very difficult to achieve for the case of the semiconducting nanowires, since their diameter of a few tens of nanometers and their length of several hundreds of nanometers do not allow for a microscopic calculation of the MBS wavefunction. Typically, the spatial profile is obtained with simple one-dimensional models~\cite{Klinovaja-PRB-2012}. The transverse profile has so far been obtained numerically for effective models: of core-shell nanowires in cylindrical~\cite{Lim-EPL-2013,Osca-EPJB-2014} and prismatic~\cite{Manolescu-PRB-2017,Stanescu-BJN-2018}, and of full nanowires in hexagonal~\cite{Woods-PRB-2018} geometries.

\begin{figure}[h!]
\includegraphics[width=\columnwidth]{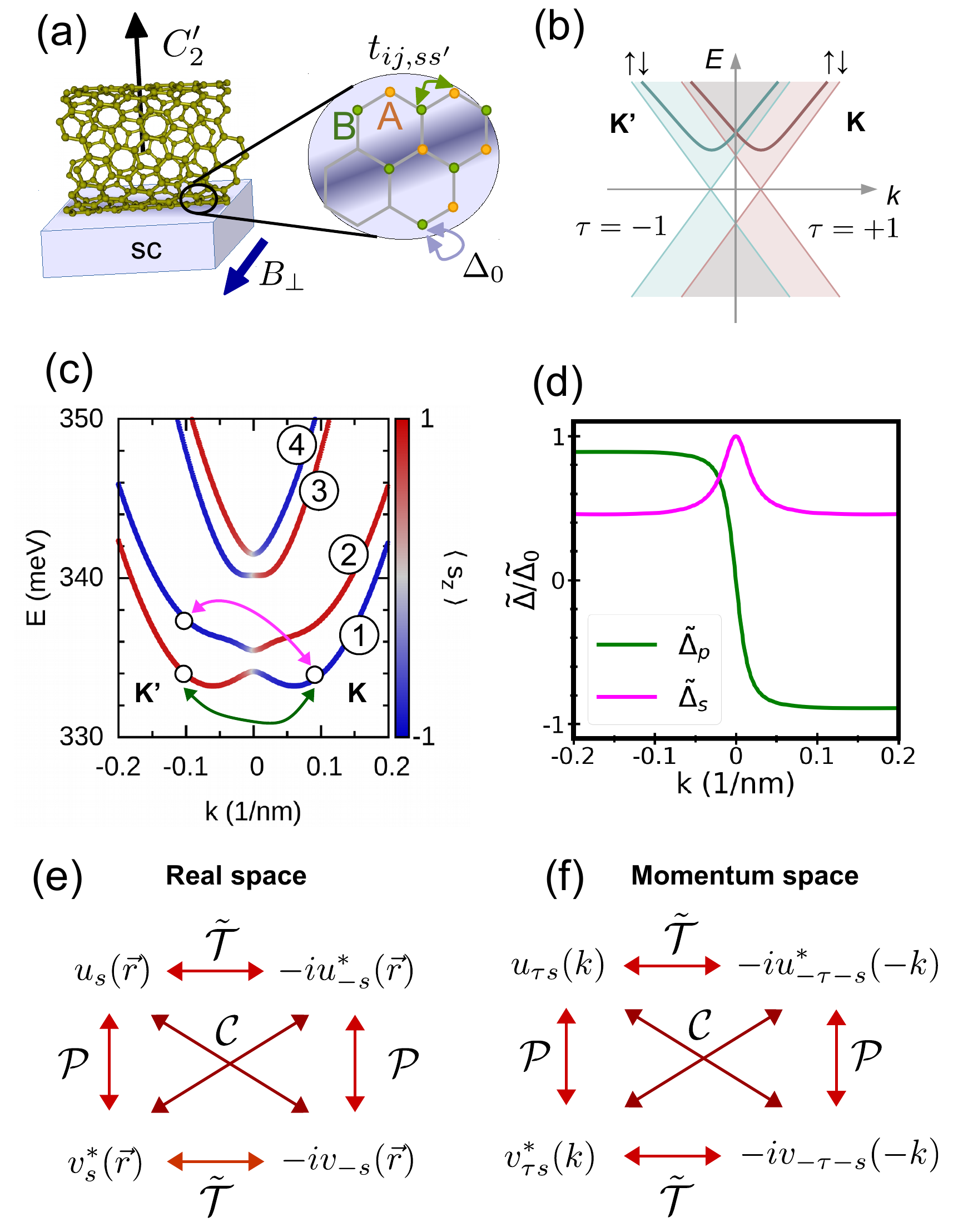}
\caption{Setup and bulk properties of a (12, 4) carbon nanotube with proximity-induced superconductivity. 
(a) Schematic of the system including the CNT which lies on top of an $s$-wave superconductor (SC) with a magnetic field applied perpendicular to the nanotube axis. The nearest neighbor hopping $t_{ij,ss^{\prime}}$ is spin-dependent due to curvature. The superconducting substrate breaks the rotational symmetry of the nanotube which induces a valley-mixing term in the Hamiltonian. Moreover it generates an on-site superconducting pairing term $\Delta_{0}$. The numerical values of the various parameters of the model can be found in Appendix \ref{sec:CNTSpecSingleParticle}.
(b) The low energy spectrum of the CNT consists of 1D cuts across the Dirac cones, with two valleys and two spin directions at each energy.
(c) The single particle energy spectrum of a (12,4) nanotube in the vicinity of the $\Gamma$-point for a magnetic field of $B_{\perp} = 14\text{T}$. Color scale shows the expectation value of $\langle s_z \rangle$ for the corresponding energy state. A finite $\Delta_{0}$ induces in the $k$-space two superconducting pairing terms $\tilde{\Delta}_{s}\left(k\right)$ and $\tilde{\Delta}_{p}\left(k\right)$ whose action is indicated by the magenta and green lines, respectively.
%$\Delta_{\mathrm{SO}}=2$~meV and $\Delta_{KK'}=2.5$~meV. 
(d) The two superconducting pairing terms $\tilde{\Delta}_{s}\left(k\right)$ (interband), and $\tilde{\Delta}_{p}\left(k\right)$ (intraband), as functions of $k$. 
(e) The action of the particle-hole $\mathcal{P}$, pseudo time-reversal $\tilde{\mathcal{T}}$ and chiral $\mathcal{C}$ operations on the components of a Nambu spinor in the real space.
(f) The counterpart of these relations in the reciprocal space. The fact that $\mathcal{P}$ relates $u_{\tau s}(k)$ and $v_{\tau s}^*(k)$ follows from $\mathcal{P}\gamma_{k}=\gamma_{-k}^\dag$.
% Note that guided by the analogy with the definition of the $\mathcal{P}$'s action on the Hamiltonian we would expect that 
% %$u_{\tau s}\left(k\right) = v_{\tau s}^{\star}\left(-k\right)$ ($u_{\tau s}\left(k\right) = -iv_{-\tau -s}\left(k\right)$). 
% it should relate $u_{\tau s}(k)$ and $v_{-\tau s}^*(-k)$. However, due to our definition of the Nambu spinor, $\mathcal{P}$ relates $u_{\tau s}(k)$ and $v_{\tau s}^*(k)$, as shown in the figure.
}
\label{fig:Setup}
\end{figure}

In this work we show that the spatial profile of MBS can be derived analytically with good accuracy in a setup which uses a carbon nanotube (CNT) in proximity with an $s$-wave superconductor.  
Similar to the nanowires, such CNTs  can host MBS at their ends \cite{Egger-PRB-2012,Klinovaja-PRL-2012,Sau-PRB-2013,Hsu-PRB-2015,Milz-PRB-2018}. 
Due to their hollow character and small diameter, CNTs of several micrometers can  be simulated numerically based on tight-binding models of carbon atoms on a rolled graphene lattice~\cite{Izumida-JPSJ-2009,Klinovaja-PRB-2011}. Such simulations  allow one to accurately evaluate the excitation spectrum and local observables. Effective single-particle low energy models can be derived which well reproduce microscopic simulations~\cite{Ando-JPSJ-2000}. \\
In a recent paper~\cite{Milz-PRB-2018} we have used a four-band and an effective one-band model to calculate the topological phase diagram and the energy spectrum of proximitized semiconducting CNTs in perpendicular magnetic field, see Fig. \ref{fig:Setup}(a), with parameters obtained from a  fit to the numerical spectra\footnote{In our tight binding model we consider one $p_{z}$ orbital per atom.}. \\
In this work we use the same models to analytically obtain the full 3D spatial profile of the Majorana wave function. 
First, we exploit our knowledge of the three symmetries of the effective BdG Hamiltonian in order to derive the relations between the four components of the Majorana spinor (see Fig.~\ref{fig:Setup}(e,f)), thus reducing the number of independent components to one.
Second, we find that the presence of two angular momentum contributions (valleys) and the spin degree of freedom results in the formation of a composite, six-piece MBS whose 3D wave function has a distinctive spiral pattern with a $C_2$ symmetry, impossible to factorize into separate transverse and longitudinal profiles. Equally non-isotropic is the spin canting angle, a quantity encoding the relative phase of the spin up and spin down particle components of the Majorana wave function. A comparison with the numerical results for the MBS of a (12,4) CNT gives us confidence in the reliability of the effective model. Our results show that while simple 1D models can capture the important low energy properties of the BdG spectrum, they might miss crucial features present in the full 3D wave function.
This can have profound implications in various setups, where the shape and local spin composition of an MBS are relevant\cite{Prada-PRB-2017,Hoffman-PRB-2017,Schuray-PRB-2018}.
 
The paper has the following structure. In Sec.~\ref{sec:setup-symmetries} we discuss our microscopic model of the carbon nanotube, the symmetries of the BdG Hamiltonian in our setup and the resulting relations between the components of the Majorana spinor. 
In Sec. \ref{sec:spin-canting} we show and discuss the numerical results of the spin canting of the full 3D MBS. We proceed to reconstruct the MBS analytically.
First we introduce in Sec. \ref{sec:BandStructure} the effective low energy model of the carbon nanotube, including the superconducting correlations. We also derive the form of the Majorana state in a continuum 1D approximation. In Sec. \ref{sec:3dMajorana} we calculate the 3D Majorana solution and determine its full spatial profile. Finally we compare the numerical results from the real-space tight-binding calculation with those of the analytical model.

\section{Model and its symmetries}
 \label{sec:setup-symmetries}

\noindent
Geometrically, a single wall carbon nanotube is equivalent to a rolled-up strip taken from the two-dimensional honeycomb of carbon atoms that makes up a graphene sheet  \cite{SaitoBook}. The band structure of the CNT can be obtained from that of graphene by imposing periodic boundary conditions in the transverse direction, which quantize the transverse momentum, turning the two-dimensional dispersion of graphene into a series of 1D cuts, which are the CNTs one-dimensional subbands, shown schematically in Fig.~\ref{fig:Setup}(b). Effective low-energy Hamiltonians can be derived from the microscopic model\cite{Ando-JPSJ-2000}. Thus, like in graphene, the low-energy band structure in nanotubes consists of two distinct and time-conjugate valleys $K$ and $K^{\prime}$ which are indexed by the quantum number $\tau$ ($\tau = +1$ for $K$ valley and $\tau =-1$ for $K^{\prime}$ valley) (cf. Fig.~\ref{fig:Setup}(b)). However, the simple fact of being rolled up drastically modifies the band structure, leading to effects that are not present in graphene. These are a curvature induced band gap and an enhanced spin-orbit coupling \cite{Ando-JPSJ-2000,Kuemmeth-Nature-2008,Izumida-JPSJ-2009,Klinovaja-PRB-2011}. The spin-orbit coupling in the nanotubes results in an effective spin-orbit field directed along the tube axis, with the sign of the field given by $\tau s$, with $s$ the spin quantum number along the CNT. The CNT's tiny diameter reduces the number of relevant transverse modes to exactly four in the low-energy regime, one for each spin and valley. 
In order to keep the low energy physics close to the $\Gamma$ point, we consider nanotubes of the zigzag class~\cite{Marganska-PRB-2015,Izumida-PRB-2016}, where the Dirac points are only slightly shifted from $k=0$.
In order to open the gap at the $\Gamma$ point, we need to remove the Kramers degeneracy between the ($\tau, s$) and ($-\tau, -s$) states. The spin degeneracy can be removed by a transverse magnetic field, but only if the valleys are also mixed. 
Fortuitously, this happens automatically when the nanotube is in contact with the bulk superconductor, i.e. the source of the proximity effect. 
Its presence breaks the rotational symmetry of the tube, introducing mixing between the $K$ and $K^{\prime}$ valley. The resulting spectrum in a normal CNT is shown in Fig.~\ref{fig:Setup}(c).

The proximity to a superconducting substrate induces Cooper pairing in the CNT. The excitation spectrum of the system can be determined from the BdG Hamiltonian, where the superconducting correlations are treated in a mean-field approximation. In the microscopic model this corresponds to an on-site pairing term \cite{Uchoa-PRL-2007}, see Fig. \ref{fig:Setup}(a), and using the Nambu spinor we can construct the microscopic BdG Hamiltonian of our system. 
To anticipate the discussion in Sec.~\ref{sec:BandStructure}, in the reciprocal space this pairing yields both an inter-band ($\tilde{\Delta}_s$, with $s$-wave symmetry) and an in-band ($\tilde{\Delta}_p$) pairing, with $p$-wave symmetry, required for topological superconductivity. The two pairings are shown in Fig.~\ref{fig:Setup}(d).

The CNT alone has a crystalline symmetry of rotation by $\pi$ around an axis perpendicular to the CNT ($C_2'$ axis in Fig.~\ref{fig:Setup}(a)). 
% The axis of this symmetry has to intersect with the CNT either in the center of an atomic bond, or in the center of a hexagon. For our system of coordinates both $x$ and $y$ axes could serve as $C_2'$ axes of an isolated CNT. The presence of substrate in the $xz$ plane forces the $C_2'$ axis to align with the $y$ direction; the presence of magnetic field applied along $x$ forces it to align with $x$.
In consequence, the CNT on superconducting substrate is a topological crystalline superconductor~\cite{Shiozaki-PRB-2014,Ando-AnnuRev-2015} with $C_2'$ axis oriented as shown in Fig.~\ref{fig:Setup}(a). In our setup, however, the $C_2'$ symmetry is broken by the magnetic field parallel to the substrate and only the local symmetries remain.\\
The true time reversal symmetry is broken by the magnetic field.
Nevertheless, the inspection of the single-particle Hamiltonian of our CNT setup in the real space~\cite{Ando-JPSJ-2000,delValle-PRB-2011,Milz-PRB-2018} shows that all its dominant terms possess a local antiunitary symmetry, which commutes with the Hamiltonian. Its action on the basis states is defined by $\tilde{\mathcal{T}} c\,\vert i s \rangle = -i c^*\,\vert i,-s\rangle$.  
Contrary to the true time reversal, $\tilde{\mathcal{T}}$ has bosonic nature $\tilde{\mathcal{T}}^{2} = 1$. The $\tilde{\mathcal{T}}$ is discussed further in the Appendix~\ref{sec:pseudoTRS}.\\
The second local symmetry is the particle-hole symmetry $\mathcal{P}$, inherent in all BdG systems. With the $\mathcal{P}$ and $\tilde{\mathcal{T}}$ symmetries combined, the BdG Hamiltonian of the nanotube is also chiral symmetric under $\mathcal{C} = \tilde{\mathcal{T}}\mathcal{P}$. When acting on the eigenstates of the finite system, expressed in the Nambu space as $\hat{\Psi}(\vec{r}) = \sum_s [u_s(\vec{r}) c_s(\vec{r}) + v_s(\vec{r}) c_s^\dag(\vec{r})]$, these operators convert between the $u_s$ and $v_s$ components of the different states in the way shown schematically in Fig.~\ref{fig:Setup}(e). (The $\tilde{\mathcal{T}}$ relation has been noticed in Ref.~\onlinecite{Hoffman-PRB-2017}, although without attributing it to the presence of a pseudo-time-reversal symmetry.) The complementary relations holding in the reciprocal space, calculated in Sec.~\ref{sec:3dMajorana}, are shown in Fig.~\ref{fig:Setup}(f). The presence of these three symmetries has a profound impact on the Majorana state. 

The wave function of the Majorana bound state is given by $\Braket{\vec{r}|\Psi_{M}} = \Psi_{M}\left(\vec{r}\right)$, where $\Ket{\Psi_{M}} = \hat{\gamma}_{M}\Ket{0}$ and $\hat{\gamma}_{M}^{\dagger} = \hat{\gamma}_{M}$ is the Majorana creation operator. Here $\vec{r} = \left(z,r_{\perp}\right)$, where $z$ and $r_{\perp}$ denote the longitudinal and the transverse components, respectively. The MBS is described by a spinor,  $\Psi_{M}\left(\vec{r}\right)=\left(u_{M\uparrow}\left(\vec{r}\right)\text{, }u_{M\downarrow}\left(\vec{r}\right)\text{, }v_{M\uparrow}\left(\vec{r}\right)\text{, }v_{M\downarrow}\left(\vec{r}\right)\right)^{T}$, with    $u_{Ms}\left(\vec{r}\right)$ and $v_{Ms}\left(\vec{r}\right)$ the electron and hole components, respectively, and  $s$ indicating the spin degree of freedom. As detailed below, it is enough to find  the $u_{M\uparrow}\left(\vec{r}\right)$ components and use the symmetries of the underlying Bogoliubov-de Gennes (BdG) Hamiltonian to determine the rest.

The first relation is a consequence of the fundamental property $\mathcal{P}\Psi_{M}\left(\vec{r}\right) \overset{!}{=} \Psi_{M}\left(\vec{r}\right)$ of a Majorana state. Thus the relation $\mathcal{P}u_{s}\left(\vec{r}\right) = v_{s}^{\star}\left(\vec{r}\right)$ becomes $u_{Ms}\left(\vec{r}\right) = v_{Ms}^{\star}\left(\vec{r}\right)$. As we will show in Section~\ref{sec:BandStructure}, the MBS are also eigenstates of the chiral symmetry $\mathcal{C}$, implying $v_{Ms}\left(\vec{r}\right) = iu_{M,-s}\left(\vec{r}\right)$. Finally, since $\mathcal{C} = \tilde{\mathcal{T}}\mathcal{P}$, the Majorana state must be an eigenstate of $\tilde{\mathcal{T}}$ as well, yielding the last relation $u_{Ms}\left(\vec{r}\right) = -iu_{M,-s}^{\star}\left(\vec{r}\right)$. The relations illustrated in Fig.~\ref{fig:Setup}(e,f) become equalities within the Majorana spinor.

\section{Spin canting of the Majorana state}
\label{sec:spin-canting}

In the nanowire/quantum dot setups where the character of the potential MBS is determined by analyzing its coupling to the discrete levels of a quantum dot, the spin canting of the MBS turns out to play an important role.~\cite{Prada-PRB-2017,Hoffman-PRB-2017,Schuray-PRB-2018} If there is a mismatch between the spin of the MBS and that of the electron on the quantum dot, the coupling is suppressed. Thus we turn next to examine the local spin canting angle in our Majorana nanotube.\\  
We first notice that the total spin of the Majorana particle, summed over both particle and hole contributions, is zero. Thus, we focus on the relative spin composition of the particle components, $(u_{M\uparrow},u_{M\downarrow})$. These are complex quantities for the considered CNT setup. 
The local expectation value for each spin direction in the particle sector is given by $\langle \vec{u}_M(\vec{r})|s_{\alpha}|\vec{u}_M(\vec{r})\rangle$, 
where $s_{\alpha}$ are the Pauli matrices, $\alpha = x,y,z$, and $\vec{u}_M\left(\vec{r}\right) = \left(u_{M\uparrow}\left(\vec{r}\right)\text{, } u_{M\downarrow}\left(\vec{r}\right)\right)^{T}$ is the electron component of the wave function.

Due to the symmetry relations, see Fig. \ref{fig:Setup}(e) and Ref.~\onlinecite{Hoffman-PRB-2017}, for the Majorana state it holds
\begin{align}
\Braket{\vec{u}_M\left(\vec{r}\right)|s_{x}|\vec{u}_M\left(\vec{r}\right)} &= -2\text{Im}\left(u_{M\uparrow}^{2}\left(\vec{r}\right)\right), \nonumber \\
%= \Braket{\vec{v}\left(\vec{r}\right)|s_{x}|\vec{v}\left(\vec{r}\right)}\text{,} \nonumber \\ 
\Braket{\vec{u}_M\left(\vec{r}\right)|s_{y}|\vec{u}_M\left(\vec{r}\right)} &= -2\text{Re}\left(u_{M\uparrow}^{2}\left(\vec{r}\right)\right), \nonumber \\
%= -\Braket{\vec{v}\left(\vec{r}\right)|s_{y}|\vec{v}\left(\vec{r}\right)}\text{,} \\
\Braket{\vec{u}_M\left(\vec{r}\right)|s_{z}|\vec{u}_M\left(\vec{r}\right)} &= 0. \nonumber
%\Braket{\vec{v}\left(\vec{r}\right)|s_{z}|\vec{v}\left(\vec{r}\right)} = 0\text{.} \nonumber
\end{align}
The expectation value $\Braket{s_{z}}$ is zero because of the pseudo time-reversal symmetry.
Knowing the values of $\langle s_x(\vec{r}) \rangle$ and $\langle s_y (\vec{r}) \rangle$ we can define a local spin direction in the plane perpendicular to the nanotube,
\begin{equation}
 \theta_{xy}(\vec{r}) = \arctan\left(\frac{\langle s_{y}(\vec{r})\rangle }{ \langle s_{x}(\vec{r})\rangle} \right)
 %= \arg(u_\uparrow^2(\vec{r})) 
 = \pi/2 -2\,\arg(u_{M\uparrow}(\vec{r})).
 \label{eq:SpinCanting}
\end{equation}

\begin{figure*}[htp]
\includegraphics[width=\textwidth]{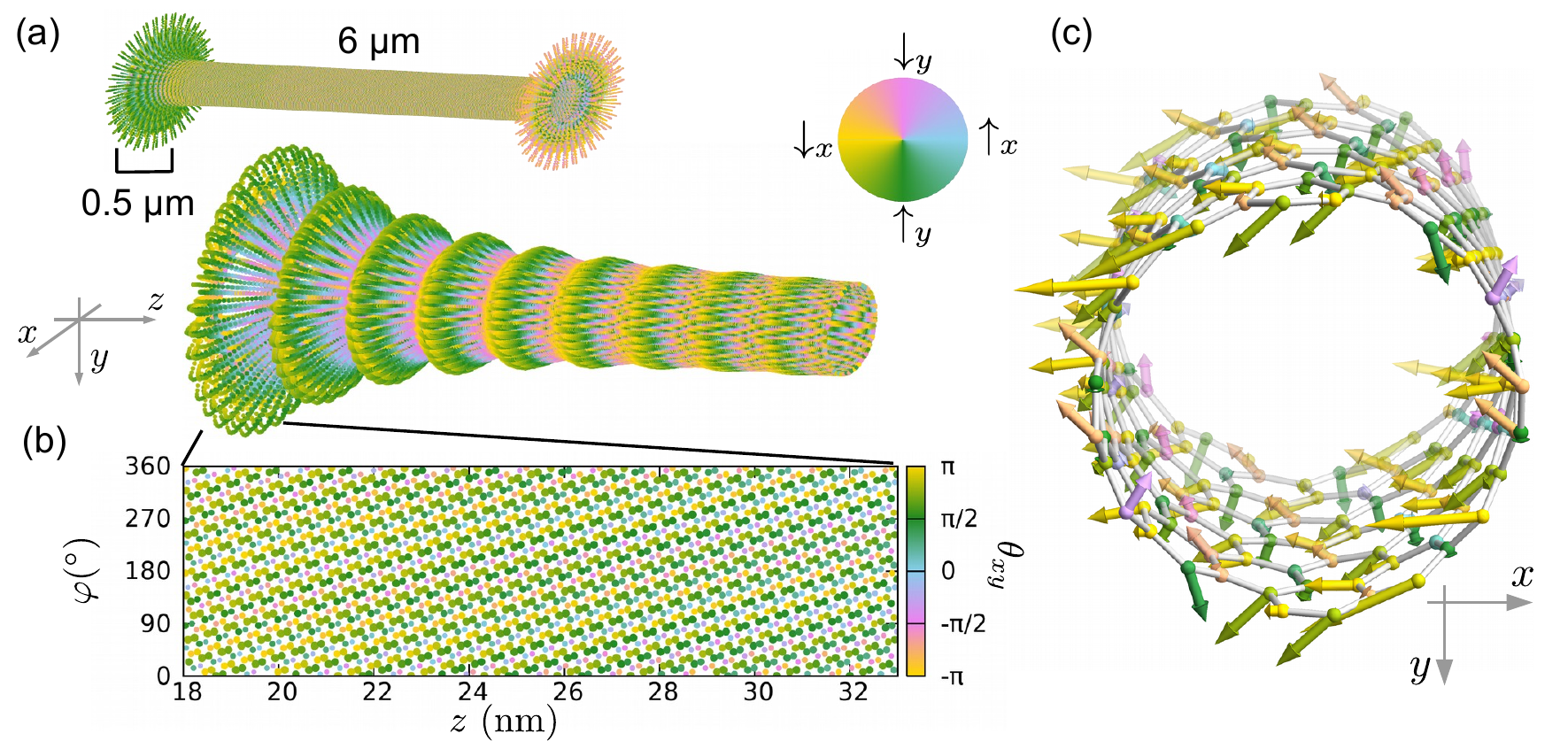}
\caption{Spin canting angle $\theta_{xy}\left(\vec{r}\right)$ and the amplitude $|u_{M\uparrow}(\mathbf{r})|$ of the electronic component of the Majorana state, obtained in a real-space tight-binding calculation of a finite (12,4) CNT with 4000 unit cells ($L = 6.03\mu m$) for a magnetic field $B_{\perp} = 14\text{T}$. 
In all panels the color corresponds to the local value of $\langle\theta_{xy}\rangle$. (a) The full Majorana state and its leftmost 0.5~$\mu$m, with distance from the CNT surface encoding $|u_{\uparrow}(\mathbf{r})|$. (b) 2D projection of the region with the first maximum of the Majorana wave function, with point size corresponding to $|u_{M\uparrow}(\mathbf{r})|$. (c) The left termination (i.e. the first 1.8 nm) of the CNT lattice. Vector length corresponds to $|u_{M\uparrow}(\mathbf{r})|$, its orientation to the spin canting angle. In both (b) and (c) note the variation of $\langle\theta_{xy}\rangle$ with the polar coordinate.}
\label{fig:SpinTexture}
\end{figure*}

The full 3D spatial profile of the wave function together with the local $\theta_{xy}(\vec{r})$ for our numerically obtained Majorana state is shown in Fig. \ref{fig:SpinTexture}(a). The distance from the CNT surface encodes the local amplitude of the MBS wave function, $|u_{M\uparrow}(\vec{r})|$, and the color scale maps $\theta_{xy}(\vec{r})$. The oscillation of $\theta_{xy}$ along $z$ with the same period as the MBS wave function is clearly visible.  Further, Fig. \ref{fig:SpinTexture}(b) shows a zoom of the left end of the tube for the first peak of $|u_{M\uparrow}(\vec{r})|$ along $z$, polar angle $\varphi$ resolved and displaying the helical pattern of $\theta_{xy}$. Finally, Fig. \ref{fig:SpinTexture}(c) visualizes the local spin canting at the very left end of the nanotube, where the electron tunneling would occur. The spin canting angle distribution takes several different values at the edge atoms, with visible $C_2$ symmetry. Thus the tunneling from a putative quantum dot coupled to the left end is definitely different than in a nanowire, assumed to be isotropic. Whether this effect is helpful or detrimental for the experiment is not yet clear.

\section{Effective four and one-band model}
\label{sec:BandStructure}

The low energy Hamiltonian of a non-superconducting CNT in the basis $\{\ket{kK\uparrow}\text{,}\ket{kK\downarrow}\text{,}\ket{kK^{\prime}\uparrow}\text{,}\ket{kK^{\prime}\downarrow}\}$ is given by
\begin{equation}
H\left(k\right) = \begin{pmatrix}
\xi_{K\uparrow}\left(k\right) & \mu_{B}B_{\perp} & \Delta_{KK^{\prime}} & 0 \\
\mu_{B}B_{\perp} & \xi_{K\downarrow}\left(k\right)  & 0 & \Delta_{KK^{\prime}} \\
\Delta_{KK^{\prime}} & 0 & \xi_{K^{\prime}\uparrow}\left(k\right) & \mu_{B}B_{\perp} \\
0 & \Delta_{KK^{\prime}} & \mu_{B}B_{\perp} &  \xi_{K^{\prime}\downarrow}\left(k\right)
\end{pmatrix}\text{,}
\label{eq:BlochCNTHam}
\end{equation}
\noindent
where $\xi_{\tau s}\left(k\right) = \varepsilon_{\tau s}\left(k\right) - \mu$ is the single-particle energy measured with respect to the chemical potential $\mu$, $\varepsilon_{\tau s}\left(k\right)$ is the single-particle energy of the electrons (see Eq. \eqref{eq:CNTdispersionFull}), $\Delta_{KK^{\prime}}$ is the energy scale associated with the valley mixing and $\mu_{B}B_{\perp}$ is the Zeeman energy due to the perpendicular magnetic field $B_{\perp}$. Diagonalization of this Hamiltonian results in four spin- and valley-mixed bands shown in Fig.~\ref{fig:Setup}(b). We can safely neglect any contributions from disorder, because CNTs can be grown with ultraclean lattices.~\cite{Cao-NatMat-2005, Deshpande-Science-2009,Jung-Nanolett-2013}
The Bloch Hamiltonian can be solved analytically with the assumption that the correlation induced by the magnetic field between lower (\ding{192},\ding{193}) and the upper (\ding{194},\ding{195}) pairs of bands is negligible~\cite{Milz-PRB-2018}. When the chemical potential is set in the lower gap at the $\Gamma$-point, this approximation allows us to consider only the lower bands $\tilde{E}_{1}\left(k\right)$ and $\tilde{E}_{2}\left(k\right)$; it holds for $\mu_BB_\perp$ smaller than both of the spin-orbit coupling and the valley mixing energy scales, which in our case are $\sim 2$~meV. The details of the calculation and a short discussion of the CNT properties is presented in the Appendix \ref{sec:CNTSpecSingleParticle}.

In the eigenbasis of \eqref{eq:BlochCNTHam} with the two-band approximation the corresponding BdG Hamiltonian for our system is given by

\begin{equation}
\tilde{\mathcal{H}}_{\text{BdG}} = \begin{pmatrix}
\tilde{E}_{1}\left(k\right) & 0 & \tilde{\Delta}_p\left(k\right) & -\tilde{\Delta}_s\left(k\right) \\
0 & \tilde{E}_{2}\left(k\right) & \tilde{\Delta}_s\left(k\right) & \tilde{\Delta}_p\left(k\right) \\
\tilde{\Delta}_p\left(k\right) & \tilde{\Delta}_s\left(k\right) & -\tilde{E}_{1}\left(k\right) & 0 \\
-\tilde{\Delta}_s\left(k\right) & \tilde{\Delta}_p\left(k\right) & 0 & -\tilde{E}_{2}\left(k\right)
\end{pmatrix}\text{.}
\label{eq:BdGHam}
\end{equation}

\noindent
Out of the two superconducting pairing terms, $\tilde{\Delta}_{s}\left(k\right) = \tilde{\Delta}_{s}\left(-k\right)$ is an even function of $k$, while $\tilde{\Delta}_{p}\left(k\right)= -\tilde{\Delta}_{p}\left(-k\right)$ is an odd function of $k$, see Fig. \ref{fig:Setup}(d). The pairing term $\tilde{\Delta}_{p}\left(k\right)$ can be viewed as a $p$-wave like gap. The BdG Hamiltonian \eqref{eq:BdGHam} can 
be partly diagonalized, taking into account the blocks with the single particle energies $\tilde{E}_{1}\left(k\right)$, $\tilde{E}_{2}\left(k\right)$ and the superconducting gap $\tilde{\Delta}_s\left(k\right)$. Details of this calculation are given in the Appendix \ref{sec:CNTPairing}. Then, the rotated BdG Hamiltonian is block-diagonal and the blocks are given by

\begin{equation}
\hat{\mathcal{H}}_{\text{BdG}}^{\pm} = \begin{pmatrix}
\tilde{\xi}_{\pm}\left(k\right) & \tilde{\Delta}_p\left(k\right) \\
\tilde{\Delta}_p\left(k\right) & -\tilde{\xi}_{\pm}\left(k\right)\\
\end{pmatrix}\text{.}
\label{eq:BdGBCSHam}
\end{equation}

\begin{figure}[h!]
\includegraphics[width=\columnwidth]{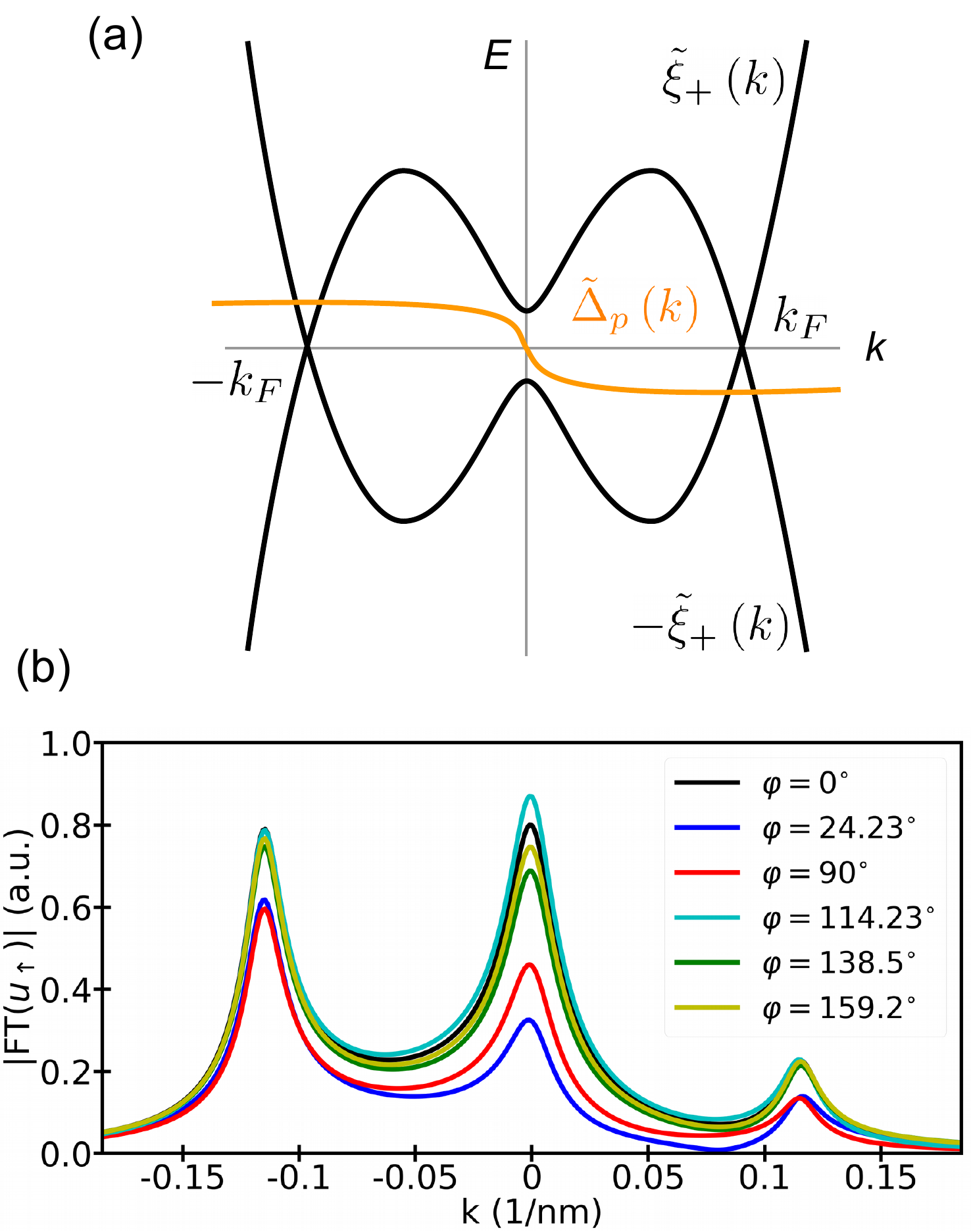}
\caption{$k$-space properties of a proximitized CNT in magnetic field at low energies. (a) Quasiparticle energy $\tilde{\xi}_{+}\left(k\right)$ and superconducting order parameter $\tilde{\Delta}_p\left(k\right)$ in the effective one-band model. 
The superconducting order paramater is an odd function of the momentum $k$. Three $k$ values generate the dominant contributions to zero energy modes: one comes from the $\Gamma$-point and one from each of the Fermi points, $\pm k_{F}$. (b) The Fourier transform of the numerical Majorana wave function for different azimuthal cuts $\varphi$ confirms that the zero mode contains only three dominant $k$ contributions.}
\label{fig:LowEnergyHam}
\end{figure}

\noindent
The quasiparticle energies $\tilde{\xi}_{\pm}\left(k\right)$ are

\begin{align*}
\tilde{\xi}_{\pm}\left(k\right) =& \frac{1}{2}\left(\tilde{E}_{1}\left(k\right) - \tilde{E}_{2}\left(k\right)\right) \\
&\pm\frac{1}{2}\sqrt{\left(\tilde{E}_{1}\left(k\right) + \tilde{E}_{2}\left(k\right)\right)^{2}+4\tilde{\Delta}_s^{2}\left(k\right)}.
\end{align*}
\noindent
The functions $\tilde{\xi}_{+}\left(k\right)$ and $\tilde{\Delta}_p\left(k\right)$ are sketched in Fig. \ref{fig:LowEnergyHam}(a).
The low energy physics, relevant for the Majorana states, is described by the block $\hat{\mathcal{H}}_{\text{BdG}}^{+}$. The particle-hole symmetry operator for the $\hat{\mathcal{H}}_{\text{BdG}}^{+}$ block is $\mathcal{P} = \tau_{x}\mathcal{K}$, and the chiral symmetry operator is $\mathcal{C} = \tau_{y}$, where $\tau_{x,y,z}$ are the Pauli matrices acting in the two-dimensional subspace of each block.

\section{Analytical reconstruction of the 3D MBS wave function}
\label{sec:3dMajorana}
\subsection{1D Majorana profile}

Majorana bound states are zero energy eigenstates of the BdG Hamiltonian and of the particle-hole symmetry operator.  From the behavior of $\tilde{\xi}_{+}\left(k\right)$ we infer that the low-energy physics has three contributions: one from the $\Gamma$-point and one from each of the Fermi points. This ansatz is confirmed by the Fourier transforms for several azimuthal cuts ($\varphi = r_\perp/R =$ const) of the numerically obtained MBS wave function, shown in Fig.~\ref{fig:LowEnergyHam}(b). 
One clearly sees one peak at the $\Gamma$-point and two peaks at opposite momenta. The peak locations are independent of $\varphi$ but their height is not. Furthermore, the peak at negative $k$ is larger. This is caused by the helical spin structure of the single-particle spectrum, shown in Fig. \ref{fig:Setup}(b). The solution at $\pm k_{F}$ is generated mostly by the band \ding{192}, and spin $\uparrow$ for this band is associated with $k < 0 $. \\
Thus, similar to some 1D models for nanowires~\cite{Klinovaja-PRB-2012}, the generic form of a Majorana state can be defined as
\begin{equation}
\ket{\Psi_{M}} = \frac{A_{\Gamma}}{\sqrt{2}}\ket{\Psi_{\Gamma}} + \frac{A_{R}}{\sqrt{2}}\ket{\Psi_{k_{F}}} + \frac{A_{L}}{\sqrt{2}}\ket{\Psi_{-k_{F}}} \text{.}
\label{eq:MajoranaState}
\end{equation}
We will later take into account the 3D nature of each of these three contributions and reconstruct the 3D spatial profile of the Majorana wave function. For now 
we approximate $\hat{\mathcal{H}}_{\text{BdG}}^{+}\approx \hat{\mathcal{H}}_{\text{BdG}}^{\Gamma}+\hat{\mathcal{H}}_{\text{BdG}}^{R}+\hat{\mathcal{H}}_{\text{BdG}}^{L}$, where we make Taylor expansions around the momenta $k = 0$ and $k = \pm k_{F}$, with $k_{F}$ determined by the constraint $\tilde{\xi}_{+}\left(k_{F}\right) = 0$.  The details of the calculation are presented in Appendix \ref{sec:MajoWave}. 

Crucially, the spinorial components of the solutions at each of the three $k$ points are the same, which allows us to combine them into a single state which is also an eigenstate of both $\mathcal{P}$ and $\mathcal{C}$. With the three contributions we can construct the 1D solution from the generic solution \eqref{eq:MajoranaState}. It is characterized by an exponential decay governed by the imaginary wave vectors $\kappa_i$ ($i=\Gamma,L,R$). The coefficients can be determined by the three constraints

\begin{subequations}
\begin{equation}
\mathcal{P}\Psi_{M}\left(z\right) \overset{!}{=} \Psi_{M}\left(z\right)\text{,} \label{eq:phs}
\end{equation}
\begin{equation}
\Psi_{M}\left(z = 0\right) \overset{!}{=} 0\text{,} \label{eq:boundary}
\end{equation}
\begin{equation}
\int_{0}^{\infty}dz\left|\Psi_{M}\left(z\right)\right|^{2}  \overset{!}{=} 1\text{.} \label{eq:normalization}
\end{equation}
\end{subequations}

\noindent
From previous findings~\cite{Milz-PRB-2018} we know that in the topological regime $\kappa_{\Gamma}\in\mathds{R}$ and $\kappa_{R}\text{, }\kappa_{L}\in\mathds{C}$. Moreover, it holds that $\text{Re}\left(\kappa_{R}\right)=\text{Re}\left(\kappa_{L}\right)$ and $\text{Im}\left(\kappa_{R}\right) = -\text{Im}\left(\kappa_{L}\right)$ $\Leftrightarrow\kappa_{R} = \kappa_{L}^{\star}$. Therefore, the wave function can be written as 

\begin{equation*}
\Psi_{M,1D}\left(z\right) = \left[\frac{A_{\Gamma}}{\sqrt{2}}e^{\kappa_{\Gamma}z} +  \frac{A_{R}}{\sqrt{2}}e^{\kappa_{R}z} +   \frac{A_{L}}{\sqrt{2}}e^{\kappa_{R}^{\star}z}\right]\begin{pmatrix}
\mp i \\
1
\end{pmatrix}\text{.}
\end{equation*}

\noindent
These eigenvectors are not eigenstates of the particle-hole operator $\mathcal{P} = \tau_{x}\mathcal{K}$, but we can multiply them by a complex number $c_{\pm} = \pm1 + i$ , such that they satify the Majorana constraint. Then, by applying the Majorana ~\eqref{eq:phs} and the boundary~\eqref{eq:boundary} conditions we get the 1D solution, which is given by 
\begin{equation}
\Psi_{M,1D}\left(z\right) = \frac{\mathcal{N}}{2}\left(\psi_{\parallel}\left(z\right)  + \psi_{\parallel}^{\star}\left(z\right) \right) \frac{1}{\sqrt{2}}\begin{pmatrix}
1 - i \\
1 + i
\end{pmatrix}\text{,}
\label{eq:1DMajoranaSolution}
\end{equation}
\noindent
where 
\begin{equation}
\label{eq:psi-parallel}
\psi_{\parallel}\left(z\right) = \left(e^{\kappa_{F}z+ik_{F}z} - e^{\kappa_{\Gamma}z}\right)
\end{equation}
encodes the dependence of the wave function on the longitudinal coordinate. The sum $\psi_\parallel(z)+\psi^*_\parallel(z)$ satisfies the boundary condition \eqref{eq:boundary}, and $\mathcal{N}$ is the normalization constant determined from \eqref{eq:normalization}. The contribution from the $\Gamma$-point is a pure evanescent state and from the contribution from the Fermi points we get a decaying oscillation with the wavevector $k_{F}$. 

\subsection{Reconstructing the 3D profile}

\noindent

In the remaining part of this work we will provide the analytical form only for $u_{\uparrow}\left(\vec{r}\right)$ (dropping the $M$ subscript for compactness of notation), since the remaining Majorana spinor components can be obtained by the application of $\mathcal{P}$, $\tilde{\mathcal{T}}$ and $\mathcal{C}$ symmetries.\\
The Majorana operator to create the state \eqref{eq:MajoranaState} is defined as
\begin{align*}
\hat{\gamma}_{M} &= \sum_{k}\begin{pmatrix}
u \\
v
\end{pmatrix}^{T}\begin{pmatrix}
d_{k+} \\
d_{-k+}^{\dagger}
\end{pmatrix}\text{,}
\end{align*}
\noindent
where $k\in\{\Gamma\text{, }k_{F}\text{, }-k_{F}\}$ and  $u = v^{\star}= \frac{1 - i}{\sqrt{2}}$. In order to find the analytical wave function we need to transform the wave function from the one-band back to the four-band model; this procedure is discussed in Appendix \ref{sec:MajoTrans}.
To express the Majorana state in the sublattice- and spin-resolved basis we  need the transformations reversing \eqref{eq:FirstTrans}, \eqref{eq:SecondTransOne} and \eqref{eq:LastTrans}. At the end we obtain
\begin{equation}
\label{eq:MajoranaValleySpin}
\hat{\gamma}_{M} = \sum_{k,\tau, s}\left(u_{\tau s}\left(k\right)c_{k\tau s} + v_{\tau s}\left(-k\right)c_{-k\tau s}^{\dagger}\right)\text{,}
\end{equation}
\noindent
for $k\in\{\Gamma,\pm k_{F}\}$, where the coefficients $u_{\tau s}\left(k\right)$ correspond to the electron and $v_{\tau s}\left(k\right)$ to the hole contribution, respectively. We find a compact form for the coefficients
\begin{align}
\label{eq:u-tau-s}
u_{\tau s}\left(k\right) = \tau s\Lambda_{\tau s}\left(k\right)\lambda_{s}\left(k\right)\text{,} \\
v_{\tau s}\left(k\right) = \tau s\Lambda_{\tau s}\left(k\right)\lambda_{s}^{\star}\left(k\right)\text{,}
\end{align}
\noindent
with
\begin{equation*}
\Lambda_{\tau s}\left(k\right) = \begin{cases}
a_{s}\left(k\right) & \text{for }\tau = +1 \text{,} \\
b_{s}\left(k\right) & \text{for }\tau = -1 \text{,}
\end{cases}
\end{equation*}
\noindent
and (see Eq.~\eqref{eq:FirstTransValues} for $a_s(k)$ and $b_s(k)$)
\begin{equation*}
\lambda_{s}\left(k\right) = u\,m\left(k\right)g\left(sk\right) - s\,v\,n\left(k\right)h\left(sk\right)\text{.}
\end{equation*}

\noindent
The coefficients $g(k),h(k)$ and $n(k),m(k)$ are found below, in Eqs.~\eqref{eq:SecondTransValues} and \eqref{eq:LastTransValues}, respectively.
By using the relations $a_{s}\left(-k\right) = b_{-s}\left(k\right)$, $g\left(-k\right) = h\left(k\right)$, $m\left(-k\right) = m\left(k\right)$, $n\left(-k\right) = n\left(k\right)$, we obtain $\Lambda_{\tau s}\left(k\right) = \Lambda_{-\tau -s}\left(-k\right)$ and $\lambda_{s}\left(k\right) = -i\lambda_{-s}^{\star}\left(-k\right)$. Finally, we arrive at the symmetry relations of the electron and hole coefficients $u_{\tau s}\left(k\right)$ and $v_{\tau s}\left(k\right)$ illustrated in Fig. \ref{fig:Setup}(f).\\

\noindent
We have now the expression of the wave function in conduction basis. In order to apply the boundary condition it must however be recast in the sublattice-resolved basis. In general for the transformation into the sublattice basis one needs also the valence band contribution. Here we can use the fact that, due to the high chemical potential, we are far away from the charge neutrality point and therefore the contribution from the valence band is negligible. With this the components in the sublattice basis are defined as $u_{p\tau s}\left(k\right) = e^{ip\eta_{\tau sk}}u_{\tau s}\left(k\right)$, where $\eta_{\tau s}\left(k\right) = \text{arg}\left(\gamma_{\tau s}\left(k\right)\right)$ is the phase of \eqref{eq:CNTKinetcEnergy} in the low-energy regime, and $p = +1$ for $A$ sublattice and $p = -1$ for $B$ sublattice.

Since our nanotube is chiral, the open boundary conditions imply that the wave function must vanish on one end at the missing $A$ atoms and on the other end at the missing $B$ atoms\cite{delValle-PRB-2011}. We use therefore the open boundary condition $\Psi_{A}\left(z = 0\text{, }r_{\perp}\right) \overset{!}{=} 0$ $\forall r_{\perp}$. The wave function  $u_{p\uparrow}\left(\vec{r}\right)$ is given by the superposition of the three contributions $k\in\{\Gamma\text{, }k_{F}\text{, }-k_{F}\}$ and the two valleys $K$ and $K^{\prime}$, each with its specific transverse profile $e^{i\tau k_\perp x_\perp}$:
\begin{widetext}
\begin{align}
u_{p\uparrow}\left(\mathbf{r}\right) &=\frac{A_{\Gamma}}{\sqrt{2}}\left[
e^{ip\eta_{K\uparrow\Gamma}}u_{K\uparrow}\left(\Gamma\right)e^{ik_{\perp}r_{\perp}}
+ e^{ip\eta_{K^{\prime}\uparrow\Gamma}}u_{K^{\prime}\uparrow}\left(\Gamma\right)e^{-ik_{\perp}r_{\perp}}\right]e^{\kappa_{\Gamma}z}\nonumber\\
&+\frac{A_{R}}{\sqrt{2}}\left[e^{ip\eta_{K\uparrow k_{F}}}u_{K\uparrow}\left(k_{F}\right)e^{ik_{\perp}r_{\perp}}+
e^{ip\eta_{K^{\prime}\uparrow k_{F}}}u_{K^{\prime}\uparrow}\left(k_{F}\right)e^{-ik_{\perp}r_{\perp}}\right]e^{\kappa_{F}z+ik_{F}z} \\\nonumber
&+\frac{A_{L}}{\sqrt{2}}\left[e^{ip\eta_{K\uparrow -k_{F}}}u_{K\uparrow}\left(-k_{F}\right)e^{ik_{\perp}r_{\perp}} +e^{ip\eta_{K^{\prime}\uparrow -k_{F}}}u_{K^{\prime}\uparrow}\left(-k_{F}\right)e^{-ik_{\perp}r_{\perp}}\right]e^{\kappa_{F}z-ik_{F}z}\text{.} \nonumber
\end{align}
\end{widetext}
\noindent
The amplitudes can be fixed by observing that
the Majorana condition requires $A_{\Gamma}\in\mathds{R}$ and $A_{R} = A_{L}^{\star}$. From the open boundary condition in longitudinal direction we obtain a relation between $A_R$ and $A_\Gamma$; hence the particle component of the wave function can be written as  
\begin{align}
u_{p\uparrow}\left(\vec{r}\right) =&\frac{1}{\sqrt{2}}\sum_{\tau}e^{i\tau k_{\perp}r_{\perp}}\bigl[A_{R}e^{ip\eta_{\tau\uparrow k_{F}}}u_{\tau\uparrow}\left(k_{F}\right)\psi_{\parallel}\left(z\right) \label{eq:WaveFunctionAUp}\\ \nonumber 
+&A_{R}^{\star}e^{ip\eta_{\tau\uparrow -k_{F}}}u_{\tau\uparrow}\left(-k_{F}\right)\psi_{\parallel}^{\star}\left(z\right)\bigr]\text{.}
\end{align}
\noindent
The expressions for $u_{\tau s}(k)$ are given in Eq.~\eqref{eq:u-tau-s}, and for $\psi_\parallel(z)$ in Eq.~\eqref{eq:psi-parallel}.
The spatial profile of the wave function is not trivial, in the sense that it cannot be factorized into separate longitudinal and transverse profiles, $u_{p\uparrow}\left(\vec{r}\right)\neq f\left(r_{\perp}\right)g\left(z\right)$. The absolute value $\vert A_R\vert$ is fixed by the normalization and its phase by the Majorana condition. 
Note that the transverse momentum $k_{\perp}$ is quantized by the periodic boundary condition. The Fermi wavevector $k_F$ is given by the position of the chemical potential $\mu$, and the characteristic decay lengths at $\Gamma$ and $\pm k_F$ by the parameters of the Hamiltonian at this $\mu$. 
Thus all factors in the wave function are in principle known from the analytics. 

\begin{figure}[h!]
\includegraphics[width=\columnwidth]{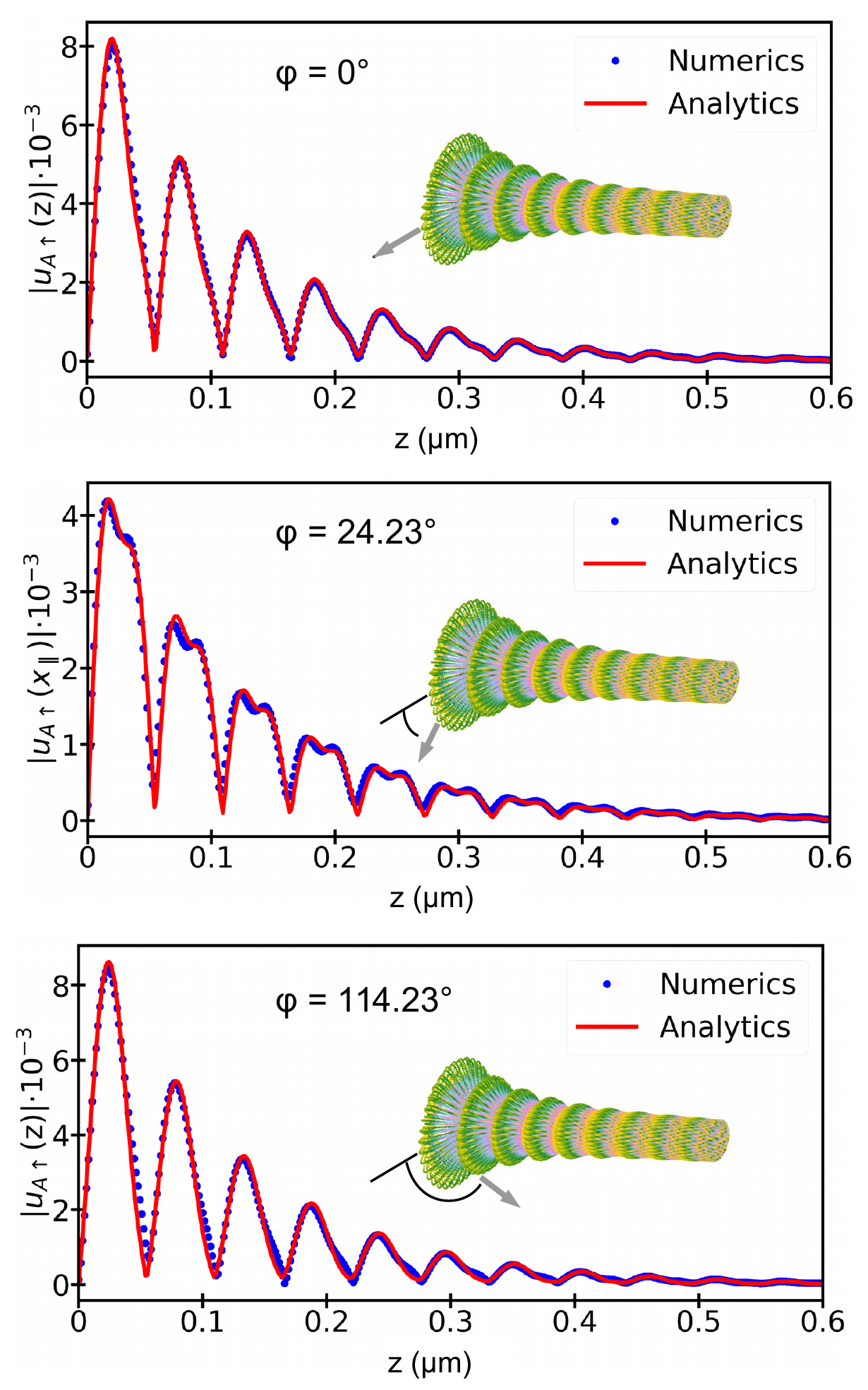}
\caption{Azimuthal cuts of the electron component $u_{A\uparrow}$ of the Majorana spinor for (a) $\varphi = 0^{\circ}$, (b) $\varphi=24.23^\circ$ and (c) $\varphi = 114.23^{\circ}$. The position of the cut in the full MBS wave function is indicated in each inset. The analytical form of $u_\uparrow$ is given by \eqref{eq:WaveFunctionAUp}, its parameters are obtained from fits to the modulus of the numerical solution.}
\label{fig:MajoranaState}
\end{figure}

\subsection{Comparison between analytical and numerical results}

\noindent
In order to test the accuracy of our formula Eq.~\eqref{eq:WaveFunctionAUp}, we have performed a comparison between the analytical and numerical solutions for several 1D cuts of the full MBS profile, at varying values of the azimuthal angle $\varphi$. We fitted the numerical solutions with \eqref{eq:WaveFunctionAUp}, finding for each cut the parameters $\kappa_\Gamma, \kappa_F, k_F$ and $A_R$. \\
The results for three values of the polar angle, $\varphi = 0^{\circ},24.23^\circ, 114.23^{\circ}$ are shown in Fig. \ref{fig:MajoranaState}. The analytical model clearly reproduces very well the numerically obtained wave functions. However, due to the simplifications inherent in the effective one-band model, there are three aspects where we have to adjust for the lost information.\\
(i)  In the microscopic model the $\mathcal{P}$ symmetry holds exactly (by construction), but $\tilde{\mathcal{T}}$ is minimally broken by two small effects. One is the presence of the weak spin-flip terms in the Hamiltonian, due to the enhanced spin-orbit coupling \cite{Ando-JPSJ-2000,Izumida-JPSJ-2009,delValle-PRB-2011}. The other is the small Peierls phase for the nearest neighbor hopping, due to the magnetic field \cite{Peierls-ZPhys-1933}. Thus in the numerical solution the $\tilde{\mathcal{T}}$- and $\mathcal{C}$-related components of the Nambu spinor differ by about $\pm 3\%$. 
Removing the spin-flip and the Peierls phase restores the $\tilde{\mathcal{T}}$ and consequently also the $\mathcal{C}$ symmetries, see Appendix~\ref{sec:pseudoTRS} for details.\\
(ii) In the analytics we neglected some correlations due to the magnetic field. Further, we performed Taylor expansions around the three momenta $k=0,\pm k_F$. Thus, the values $\kappa_{\Gamma}$, $\kappa_{F}$ and $k_{F}$ from the analytics are slightly different from those which are obtained by fitting the numerical data using \eqref{eq:WaveFunctionAUp}, see Tab. \ref{tab:tab1}.

\begin{table}
\begin{tabular}{c|c|c}
& Analytics ($\frac{1}{\mu m}$) & Fits ($\frac{1}{\mu m}$)\\
\hline
$\kappa_{\Gamma}$ &  -7.94 & -8.93 \\
$\kappa_{F}$ & -6.56 & -8.01 \\
$k_{F}$ & 118.92 & 115.25 
\end{tabular}
\caption{Values of $k_{F}$, $\kappa_{\Gamma}$ and $\kappa_{F}$ from the analytical calculation compared with values fitted from the numerics.}
\label{tab:tab1}
\end{table}

\begin{figure}[h]
\includegraphics[width=1.0\linewidth]{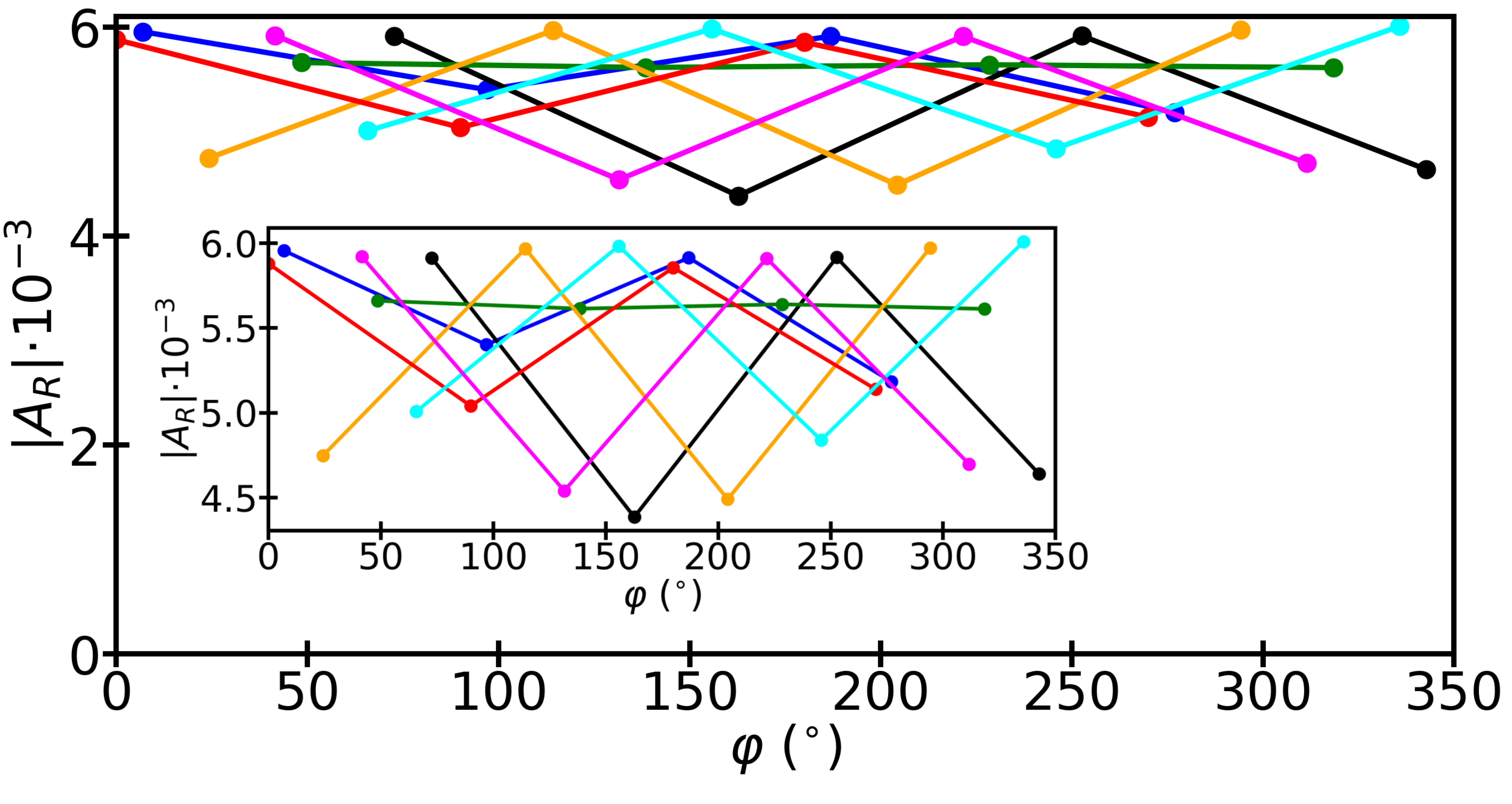}
\caption{The absolute value of the fitted amplitude $|A_{R}|$ of 28 different $\varphi$ cuts. The colors correspond to different groups of atoms related by the $C_{4}$ symmetry (i.e. atoms at the same $z$ position). From the inset we see the approximate $\pi$-periodicity of $A_R$ and thus the $C_{2}$ symmetry of the MBS wave function.}
\label{fig:Amplitude}
\end{figure}

(iii) We implemented the valley mixing through a continuous potential ridge along the CNT/superconductor interface. This results in the coupling between the two valleys, but also in their coupling to higher transverse momentum bands which therefore also contribute, albeit very weakly, to the final Majorana state. In consequence, although we expect $A_R$ to be independent of $\varphi$, we obtain from the fitting procedure different $A_R$ for different $\varphi$ cuts, with the resulting values of $|A_R(\varphi)|$ shown in Fig.~\ref{fig:Amplitude}. 
%We fit 28 different cuts of $\varphi$. In particular,  $A_{R}(\varphi=0^\circ) = \left(4.358 + 3.947i\right)\cdot 10^{-3}$ and  $A_{R}(\varphi=114.23^\circ) = \left(5.009 + 3.245i\right)\cdot 10^{-3}$. 
We see that, although not constant, the amplitude $A_{R}$ is a weakly varying function of $\varphi$. Moreover, the data resolved for atoms at the same $z$ position show that $A_R$ is close to $\pi$-periodic. This is a consequence of the $C_4$ symmetry of our (12,4) CNT where the $K'/K$ valley states carry the angular momentum $\ell = \pm 1$. Since the Majorana state is constructed predominantly from electron (and hole) states with $\ell=\pm 1$, the {\em amplitude} of its wave function, to which $A_R$ was fitted, has an approximate $C_{2}$ symmetry. This is also visible in Fig.~\ref{fig:SpinTexture}(c), where the $C_2$ (instead of $C_1$) symmetry of spin texture arises from the factor of 2 in Eq.~\eqref{eq:SpinCanting}. 

%However, the Amplitude the assumption of a global constant $A_{R}$ is fine because the range of $A_{R}\left(\varphi\right)$ is from $4.0\cdot 10^{-3}$ to $6.5\cdot 10^{-3}$. Our (12,4) CNT has $C_{4}$ symmetry and the Dirac subbands carry the angular momentum $\ell=\pm 2$.

\begin{figure}[h!]
\includegraphics[width=1.0\columnwidth]{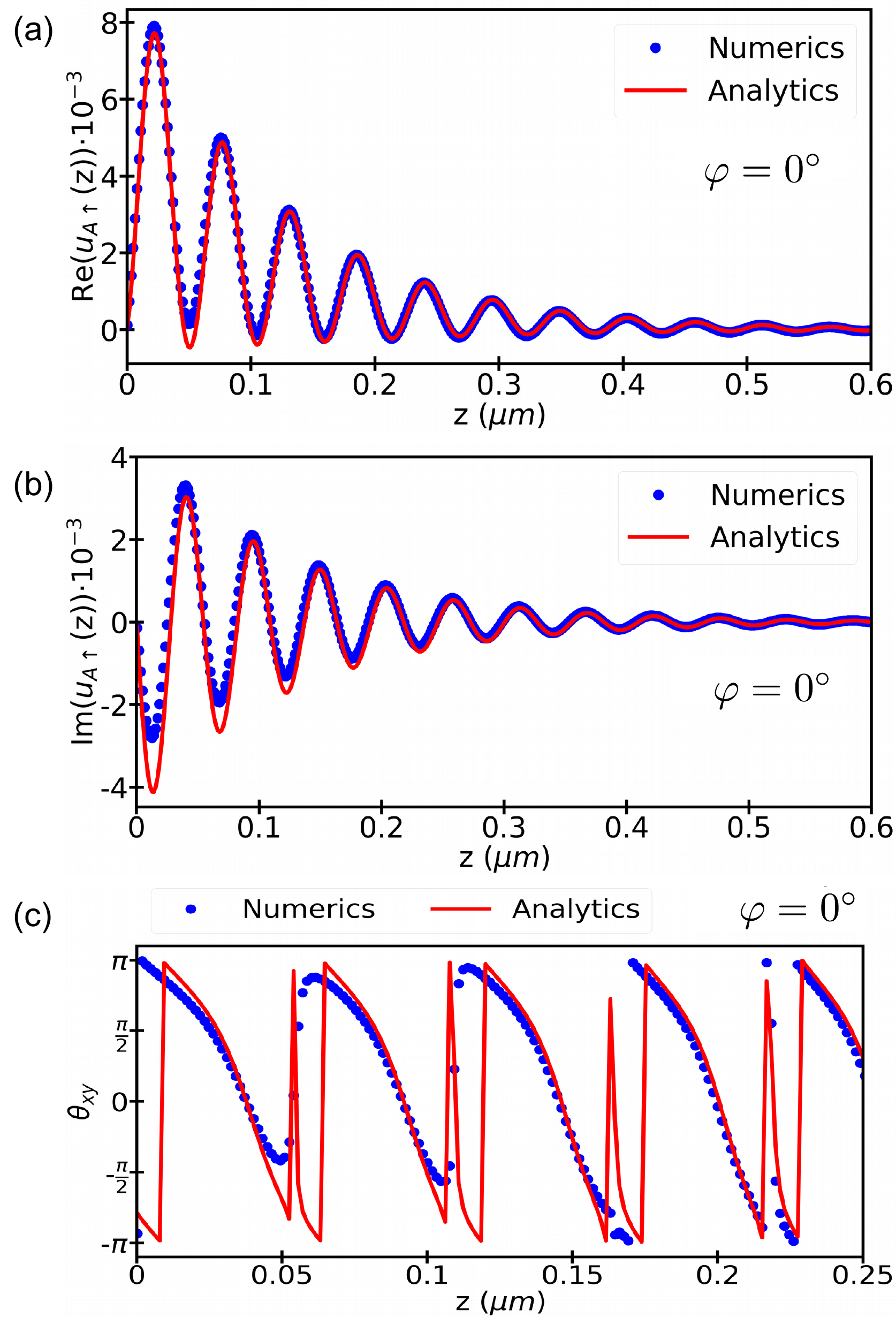}
\caption{(a)-(b) Numerical and analytical $\text{Re}\left(u_{A\uparrow}\left(z\right)\right)$ and $\text{Im}\left(u_{A\uparrow}\left(z\right)\right)$ for the polar angle $\varphi = 0^{\circ}$ with parameters from the $\left|u_{A\uparrow}\left(z\right)\right|$ fit. (c) The spin canting angle $\theta_{xy}$,  defined in Eq. \eqref{eq:SpinCanting},for the cut $\varphi = 0^{\circ}$.}
\label{fig:SpinTexture_Appendix}
\end{figure}

\noindent
In Fig. \ref{fig:SpinTexture_Appendix} we show a comparison between the analytical and numerical results for $\mathrm{Re}(u_{A\uparrow}),\mathrm{Im}(u_{A\uparrow})$ and the resulting canting angle $\theta_{xy}(z)$ for $\varphi=0$. The slight discrepancy between the numerical and analytical values of the real and imaginary part of $u_\uparrow(\vec{r})$, shown in Fig.~\ref{fig:SpinTexture_Appendix}(a-b), is amplified in the spin canting angle behavior shown in Fig.~\ref{fig:SpinTexture_Appendix}(c).
In particular, additional phase jumps are visible at positions where the real value in numerics is small and positive, while the analytical result is also small but negative.  Nevertheless, the overall agreement is again good.

\section*{Conclusion}
In this work we have shown in a combination of  numerical modelling and analytical calculations how to determine the full spatial profile of the Majorana bound state in a proximitized semiconducting carbon nanotube. The wave function has three contributions: one from the $\Gamma$-point and one from each Fermi point, which is also supported by an analysis of the numerical data via a Fourier transformation. 
We find the symmetry relations which must be fulfilled by the components of the Majorana spinor.
The excellent agreement between the analytically obtained  and the numerically calculated spin and sublattice resolved spinor gives us confidence in the accuracy  of the local observables further derived in this work. 
 Despite being obtained for a CNT, our results might serve as a reference also for other systems where a microscopic calculation of the MBS spinor is not possible.
The features which our model captures very well are: the three main momentum contributions to the MBS, the decaying behaviour of the wave function combined with its spiral pattern, its oscillation and the symmetries linking the different components of the Nambu spinor.
We show that our analytical model fits very well the numerical data of the wave function obtained by a tight-binding calculation. Our results will be useful for modeling and interpreting the experimental results in a realistic quantum transport setup where the properties of the Majorana states are probed locally.

\acknowledgments{
The authors thank the Deutsche Forschungsgemeinschaft for financial support via 
GRK 1570 and IGK ``Topological insulators'' grants, as well as the JSPS for the KAKENHI Grants (Nos. JP15K05118, JP15KK0147, JP18H04282).}

%\newpage

\appendix

\section{CNT Spectrum}

\subsection{Single-particle spectrum}
\label{sec:CNTSpecSingleParticle}

\noindent
The simplest way of obtaining the Hamiltonian of a CNT in the momentum space representation is to use the zone folding approximation\cite{SaitoBook}. The Hamiltonian of a CNT can be written in the sublattice basis for $A$ and $B$ sublattice

\begin{equation}
H_{0} = \sum_{\mathbf{k},s}\gamma_{s}\left(\mathbf{k}\right)a_{\mathbf{k}s}^{\dagger}b_{\mathbf{k}s} + \text{h.c.},
\label{eq:CNTSublattice}
\end{equation}

\noindent
where $a_{\mathbf{k}s}^{\dagger}$ ($b_{\mathbf{k}s}^{\dagger}$) creates an electron on $A$ ($B$) sublattice with momentum $\mathbf{k}$ and spin $s$. The kinetic energy $\gamma\left(\mathbf{k}\right)$ is defined as

\begin{equation}
\gamma_{s}\left(\mathbf{k}\right) = t_{s,1}e^{i\mathbf{k}\cdot\mathbf{a}_{1}} + t_{s,2} e^{i\mathbf{k}\cdot\mathbf{a}_{2}} +t_{s,3},
\label{eq:CNTKinetcEnergy}
\end{equation}

\noindent
where $t_{s,i}$ is the spin-dependent hopping parameter between an $A$ atom and its $i$-th neighbor, and $\mathbf{a}_{1}$ and $\mathbf{a}_{2}$ are the Bravais lattice vectors of the graphene lattice, see Fig. \ref{fig:Setup}(a).  The low-energy unperturbed CNT Hamiltonian $H_{\text{CNT}}$ can be obtained by an expansion of \eqref{eq:CNTSublattice} around the Dirac points $\mathbf{k} = \bm{\kappa} + \tau\mathbf{K}$\cite{delValle-PRB-2011} and a rotation from sublattice into conduction/valence band basis. In the following we will assume that the chemical potential is in the conduction band, obtaining

\begin{equation}
H_{\text{CNT}} - \mu N = \sum_{k,\tau,s}\xi_{\tau s}\left(k\right)c_{k\tau s}^{\dagger}c_{k\tau s}\text{,}
\end{equation}

\noindent
where $\xi_{\tau s}\left(k\right) = \varepsilon_{\tau s}\left(k\right) - \mu$, $\varepsilon_{\tau s}\left(k\right) = \left|\gamma_{\tau s}\left(k\right)\right|$ is the CNT single-particle energy in the conduction band, $\mu$ the chemical potential and $c_{k\tau s}^{\dagger}\Ket{0} = \Ket{k\tau s}$ define the basis of \eqref{eq:BlochCNTHam}. The curvature of the CNT's lattice results in both spin-dependent and spin-independent modifications, i.e. shifts in both transverse and longitudinal momentum. Thus, the single-particle energies of a CNT \eqref{eq:BlochCNTHam} for given transverse momentum $k_\perp$ and longitudinal momentum $k$ at low energies are given by

\begin{eqnarray}
 \varepsilon_{\tau s}\left(k_\perp,k\right) & = & \hbar v_{F} \left\{\left(k -\tau K_\parallel + \tau\Delta k_\parallel^{c}\right)^{2}\right.   \label{eq:CNTdispersionFull}
\\
& &+\left. \left(k_{\perp} -\tau K_\perp + \tau\Delta k_{\perp}^{c} + s\Delta k_{\perp}^{\mathrm{SO}}\right)^{2}\right\}^{1/2},\nonumber
\end{eqnarray}
 
\noindent
where $K_\perp,K_\parallel$ are the transverse and longitudinal components of momentum at the Dirac point $K$. The quantum numbers $\tau$ and $s$ are are defined in the main text. In the case of the (12,4) semiconducting nanotube, the numerical values of those  momentum shifts in our calculations are $\Delta k_\perp^c = -22.83\,\mu\mathrm{m}^{-1}$, $\Delta k_\parallel^c = 66.62\,\mu\mathrm{m}^{-1}$, $\Delta k_{\mathrm{SO}}=-2.917\,\mu\mathrm{m}^{-1}$, $K_\parallel=0$, and the lowest energy subbands shown in Fig. \ref{fig:Setup}(b) have $k_\perp - \tau K_\perp = \tau/3R$. The value of $k_\perp$ for the $K$ valley subband in our nanotube is $-35/R$, where $R$ is the CNT radius.
%The spin-orbit splitting $\Delta_{\mathrm{SO}}$ from the main text is then $\Delta_{\mathrm{SO}}=\varepsilon_{K\uparrow}(0)-\varepsilon_{K\downarrow}(0)$. 
Note that the single-particle energies satisfy the time-reversal conjugation, $\varepsilon_{\tau s}\left(k\right) = \varepsilon_{-\tau-s}\left(-k\right)$.

The low-energy Bloch Hamiltonian \eqref{eq:BlochCNTHam} contains also the valley mixing and Zeeman field contributions, $H = H_{\text{CNT}} + H_{\Delta_{KK^{\prime}}} + H_{Z}$. 
Since the nanotube we are studying is of the zigzag class~\cite{Marganska-PRB-2015,Izumida-PRB-2016}, in order to mix the valleys it is enough to break only the rotational symmetry.
In our setup we consider the valley mixing introduced by the presence of the substrate, modelling it as an electrostatic potential with a Gaussian distribution in the polar coordinate, $V(\varphi)=V_0 \exp(-(\varphi-\pi/2)/\Delta\varphi)$. This corresponds to the substrate extending in the $xz$ plane. In the reciprocal space the valley-mixing term is given by

\begin{equation}
H_{KK^{\prime}} = \sum_{k,s}\Delta_{KK^{\prime}}\left(c_{kKs}^{\dagger}c_{kK^{\prime}s} + c_{kK^{\prime}s}^{\dagger}c_{kKs}\right)\text{,}
\end{equation}

\noindent
and couples states with the same spin and $k$ but opposite valley. Following Ref.~\onlinecite{Milz-PRB-2018}, we set $\Delta_{KK'} = 2.5$~meV.\\
The Zeeman effect with the field aplied along the $x$ axis couples opposite spins in the same valley,

\begin{equation}
H_{Z} = \sum_{k,\tau}\mu_{B}B_{\perp}\left(c_{k\tau\uparrow}^{\dagger}c_{k\tau\downarrow} + c_{k\tau\downarrow}^{\dagger}c_{k\tau\uparrow}\right)\text{.}
\end{equation}

The CNT Hamiltonian \eqref{eq:BlochCNTHam} can be brought to a diagonal form by employing two unitary transformations. More details about the transformations can be found in Appendix D.1 of Ref. [\onlinecite{Milz-PRB-2018}]. The first transformation diagonalizes the Hamiltonian without Zeeman energy ($B_{\perp} = 0$) and is defined as
\begin{equation}
\begin{pmatrix}
c_{k Ks} \\
c_{k K^{\prime}s}
\end{pmatrix} = \begin{pmatrix}
a_{s}\left(k\right) &  b_{s}\left(k\right) \\
-b_{s}\left(k\right)& a_{s}\left(k\right)
\end{pmatrix}\begin{pmatrix}
\alpha_{k s} \\
\beta_{k s}
\end{pmatrix}\text{,}
\label{eq:FirstTrans}
\end{equation}
with $a_{s}\left(k\right)^{2} +b_{s}\left(k\right)^{2}  = 1$ and the following values of  $a_{s}\left(k\right)$ and $b_{s}\left(k\right)$,

\begin{subequations}
\label{eq:FirstTransValues}
\begin{align}
a_{s}^{2}\left(k\right)& =\frac{1}{2}\left(1 - \frac{\xi_{Ks}\left(k\right) - \xi_{K^{\prime}s}\left(k\right)}{\sqrt{\left(\xi_{K s}\left(k\right) - \xi_{K^{\prime}s}\left(k\right)\right)^{2}+4\Delta_{KK^{\prime}}^{2}}}\right)\text{,} \\
b_{s}^{2}\left(k\right) &=\frac{1}{2}\left(1 + \frac{\xi_{Ks}\left(k\right) - \xi_{K^{\prime}s}\left(k\right)}{\sqrt{\left(\xi_{Ks}\left(k\right) - \xi_{K^{\prime}s}\left(ks\right)\right)^{2}+4\Delta_{KK^{\prime}}^{2}}}\right)\text{,}
\end{align}
\end{subequations}

where the energy eigenvalues are  

\begin{align}
E_{\pm s}\left(k\right) = &\frac{1}{2}\left(\xi_{K s}\left(k\right) + \xi_{K^{\prime} s}\left(k\right)\right) \\
&\pm \frac{1}{2}\sqrt{\left(\xi_{K s}\left(k\right) - \xi_{K^{\prime} s}\left(k\right)\right)^{2}+4\Delta_{KK^{\prime}}^{2}}\text{.}\nonumber
\end{align}
Due to the time-reversal conjugation of  $\xi_{\tau s}\left(k\right) = \xi_{-\tau- s}\left(-k\right)$, it can be shown that $a_{ s}\left(k\right) = b_{- s}\left(-k\right)$ and $E_{\pm s}\left(k\right) = E_{\pm- s}\left(-k\right)$.

Using equations \eqref{eq:FirstTrans} the Zeeman term can be expressed as

 \begin{align}
 \tilde{B}_{\perp} &= B_{\perp}\left(\left|a_{\uparrow}\left(k\right)\right|\left|a_{\downarrow}\left(k\right)\right| + \left|b_{\uparrow}\left(k\right)\right|\left|b_{\downarrow}\left(k\right)\right| \right),\\
 B_{\perp}^{\star} &= B_{\perp}\left(\left|a_{\uparrow}\left(k\right)\right|\left|b_{\downarrow}\left(k\right)\right| - \left|b_{\uparrow}\left(k\right)\right|\left|a_{\downarrow}\left(k\right)\right| \right).
 \label{eq:Bfields}
 \end{align}

\noindent
The magnetic field $\tilde{B}_{\perp}$ couples the spins within the lower and upper band pair, while  $B_{\perp}^{\star}$ couples the spins between band pairs. Both are symmetric in $k$, i.e. $\tilde{B}_{\perp}\left(k\right) = \tilde{B}_{\perp}\left(-k\right)$ and $B_{\perp}^\star\left(k\right) = B_{\perp}^\star\left(-k\right)$. This is a consequence of the pseudo-time reversal symmetry. \\
In the regime of small Zeeman energy, i.e. $\Delta E = \left|E_{+ s} - E_{-  s}\right| > \mu_{B}B_{\perp}$, the terms with $B_{\perp}^{\star}$ can be omitted. This allows us to treat the upper and lower pair of bands separately. We shall proceed to find the solutions for the lower band pair only, assuming that the chemical potential $\mu$ is tuned into the gap between the two energy bands $\tilde{E}_{1}$ and $\tilde{E}_{2}$. Therefore, we will neglect the influence of the bands $\tilde{E}_{3}$ and $\tilde{E}_{4}$ because those bands are not occupied. Then, the second transformation diagonalizing the Hamiltonian with magnetic field is defined as
\begin{equation}
\begin{pmatrix}
\alpha_{k\uparrow} \\
\alpha_{k\downarrow}
\end{pmatrix} = \begin{pmatrix}
g\left(k\right) &  h\left(k\right) \\
-h\left(k\right)& g\left(k\right)
\end{pmatrix}\begin{pmatrix}
f_{k1} \\
f_{k2}
\end{pmatrix}\text{,}
\label{eq:SecondTransOne}
\end{equation}
where the coefficients must satify $g^{2}\left(h\right) + h^{2}\left(k\right) = 1$. The new quantum number in \eqref{eq:SecondTransOne}  $i\in\{1\text{,}2\}$ just reflects the ordering of the energy bands $E_{1} < E_{2}$. The coefficients $g$ and $h$ are defined as
\begin{subequations}
\label{eq:SecondTransValues} 
\begin{align}
g^{2}\left(k\right) &=\frac{1}{2}\left(1 - \frac{E_{-\uparrow}\left(k\right) - E_{-\downarrow}\left(k\right)}{\sqrt{\left(E_{-\uparrow}\left(k\right) - E_{-\downarrow}\left(k\right)\right)^{2}+4\left(\mu_{B}\tilde{B}_{\perp}\right)^{2}}}\right)\text{,}&\\
h^{2}\left(k\right) &=\frac{1}{2}\left(1 + \frac{E_{-\uparrow}\left(k\right) - E_{-\downarrow}\left(k\right)}{\sqrt{\left(E_{-\uparrow}\left(k\right) - E_{-\downarrow}\left(k\right)\right)^{2}+4\left(\mu_{B}\tilde{B}_{\perp}\right)^{2}}}\right)\text{.}
\end{align}
\end{subequations}
The coefficients satisfy the pseudo-time-reversal conjugation $g\left(k\right) = h\left(-k\right)$. Then, the single-particle energies of the full Hamiltonian with decoupled band pairs are
\begin{align}
\tilde{E}_{i}\left(k\right) = & \frac{1}{2}\left(E_{-\uparrow}\left(k\right) + E_{-\downarrow}\left(k\right)\right) + \\
&+ (-1)^i \frac{1}{2}\sqrt{\left(E_{-\uparrow}\left(k\right) - E_{-\downarrow}\left(k\right)\right)^{2}+4\left(\mu_{B}\tilde{B}_{\perp}\right)^{2}}\text{.}\nonumber
\end{align}
The renormalized magnetic field opens a band gap at the $\Gamma$-point.
The single-particle energies have the property $\tilde{E}_{i}\left(k\right) = \tilde{E}_{i}\left(-k\right)$ with $i\in\{1\text{,}2\}$ because $\tilde{B}_{\perp}\left(k\right) = \tilde{B}_{\perp}\left(-k\right)$. This pseudo-time reversal symmetry for conduction band states results in the relation depicted in Fig.~\ref{fig:Setup}(f). Since the single-particle states of a finite CNT in our setup contain both $\vec{k},s$ and $-\vec{k},-s$ contributions with equal weights, their spin components in the real space must also obey the relation shown in Fig.~\ref{fig:Setup}(e). 

\subsection{Pseudo-time reversal symmetry}
\label{sec:pseudoTRS}

\begin{figure}[h]
\begin{center}
 \includegraphics[width=0.8\columnwidth]{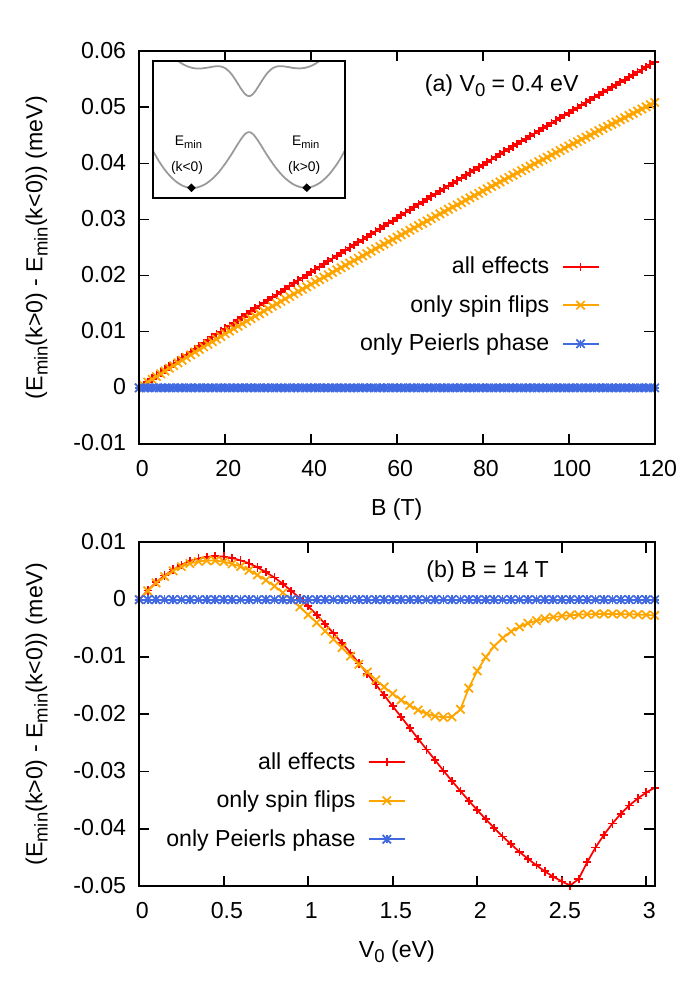}
 \end{center}
 \caption{\label{fig:pseudoTRS}
 The breaking of $\tilde{\mathcal{T}}$ as a function of $B$ and of $V_0$. 
 (a) The substrate potential has the amplitude $V_0=0.4$~eV, the value taken in the simulations shown in the main text.
 (b) The magnetic field is $B = 14$~T as in the main text and the strength $V_0$ of the substrate potential is varied. 
 Both plots show a restoration of $\tilde{\mathcal{T}}$ when the spin flips are neglected.}
\end{figure}

The pseudo-time reversal invariance holds exactly for our effective model Hamiltonian ~\eqref{eq:BlochCNTHam}. For the real space Hamiltonian it is however broken by  
two effects, both absent in our four-band model. The first and smaller one is the presence of the Peierls phase.\cite{Peierls-ZPhys-1933} This phase can be safely neglected - a magnetic field of 120~T would result in only $10^{-3}$ of a flux quantum per each hexagonal plaquette.\\
The second and more important effect is the presence of nearest-neighbor hoppings with spin flip, which couple neighboring angular momentum subbands.\cite{Ando-JPSJ-2000,Izumida-JPSJ-2009,delValle-PRB-2011} Including it would require bringing the number of subbands up to twelve.
Figure~\ref{fig:pseudoTRS} shows the strength of $\tilde{\mathcal{T}}$ breaking, quantified as the difference between the band \ding{192} (cf. Fig.~\ref{fig:Setup}) minima at $k>0$ and at $k<0$, as a function of $B$ and of $V_0$, as shown in the inset of Fig~\ref{fig:pseudoTRS}(a). 
Although the Peierls phase does contribute to the breaking of $\tilde{\mathcal{T}}$ when the spin flips are included, we see that neglecting the spin-fliping hoppings restores $\tilde{\mathcal{T}}$ completely. Nevertheless, even when all effects are present, for our parameters $B=14$~T and $V_0=0.4$~eV the pseudo-time reversal still holds down to $\mu$eV energy scales. 

\subsection{Superconducting spectrum}
\label{sec:CNTPairing}

\noindent
Including superconducting correlations on a mean-field level we add to the Hamiltonian a superconducting pairing term \cite{Uchoa-PRL-2007}, which is given by

\begin{equation}
H_{\text{SC}} = \sum_{k}\Delta_{0}\left(c_{kK\uparrow}^{\dagger}c_{-kK^{\prime}\downarrow} + c_{kK^{\prime}\uparrow}^{\dagger}c_{-kK\downarrow} + \text{h.c.}\right)\text{,}
\label{eq:PairingTerm}
\end{equation}

\noindent
where $\Delta_{0}$ is the superconducting order parameter, which we take to be 0.4~meV. We can express the pairing Hamiltonian \eqref{eq:PairingTerm} in the eigenbasis of the CNT \eqref{eq:BlochCNTHam} and, after applying the approximations and transformations described in Appendix \ref{sec:CNTSpecSingleParticle}, we obtain the BdG Hamiltonian \eqref{eq:BdGHam} with the pairing terms
\begin{align}
\tilde{\Delta}_p\left(k\right) &= \Delta_{0}\left(g^{2}\left(k\right) - h^{2}\left(k\right)\right) = -\tilde{\Delta}_p\left(-k\right)\text{,}&\\
\tilde{\Delta}_s\left(k\right) &= 2\Delta_{0}g\left(k\right)h\left(k\right) = \tilde{\Delta}_s\left(-k\right)\text{.}
\end{align}
%\noindent
We see that the pairing term $\tilde{\Delta}_s\left(k\right)$ has an even and $\tilde{\Delta}_p\left(k\right)$ an odd parity, as shown in Fig. \ref{fig:Setup}(c).\\
The basis change which transforms \eqref{eq:BdGHam} into \eqref{eq:BdGBCSHam} is given by

\begin{equation}
\begin{pmatrix}
f_{k 1} \\
f_{-k 2}^{\dagger}
\end{pmatrix}=\begin{pmatrix}
m\left(k\right) & n\left(k\right) \\
-n\left(k\right) & m\left(k\right)
\end{pmatrix}\begin{pmatrix}
d_{k+} \\
d_{-k-}^{\dagger}
\end{pmatrix}\text{,}
\label{eq:LastTrans}
\end{equation}

\noindent
with the normalization condition $m^{2}\left(k\right) + n^{2}\left(k\right) = 1$ and the coefficients defined in the following way:

\begin{subequations}
\label{eq:LastTransValues} 
\begin{align}
m^{2}\left(k\right) &= \frac{1}{2}\left(1 + \frac{E_{1}\left(k\right) + E_{2}\left(k\right)}{\sqrt{\left(E_{1}\left(k\right) + E_{2}\left(k\right)\right)^{2} + \left(2\tilde{\Delta}_s\left(k\right)\right)^{2}}}\right)\text{,}&\\
n^{2}\left(k\right) &= \frac{1}{2}\left(1 - \frac{E_{1}\left(k\right) + E_{2}\left(k\right)}{\sqrt{\left(E_{1}\left(k\right) + E_{2}\left(k\right)\right)^{2} + \left(2\tilde{\Delta}_s\left(k\right)\right)^{2}}}\right)\text{.}
\end{align}
\end{subequations}

\section{1D Majorana bound state solutions}
\label{sec:MajoWave}

\noindent
Majorana bound states are zero energy eigenstates of the BdG Hamiltonian and the particle-hole symmetry operator. The low-energy physics of the BdG Hamiltonian \eqref{eq:BdGBCSHam} is described by the block $\hat{\mathcal{H}}_{\text{BdG}}^{+}$. For the Majorana bound states we will approximate the BdG Hamiltonian by $\hat{\mathcal{H}}_{\text{BdG}}^{+}\approx \hat{\mathcal{H}}_{\text{BdG}}^{\Gamma}+\hat{\mathcal{H}}_{\text{BdG}}^{R}+\hat{\mathcal{H}}_{\text{BdG}}^{L}$ because the low-energy physics of $\hat{\mathcal{H}}_{\text{BdG}}^{+}$ has three contributions, as illustrated in Fig. \ref{fig:LowEnergyHam}(a). This is also supported by the numerics, see Fig. \ref{fig:LowEnergyHam}(b).

\subsection{$\Gamma$-point contribution}

\noindent
The first contribution is coming from the $\Gamma$-point. Therefore, we obtain from a Taylor expansion around the $\Gamma$-point

\begin{align*}
\tilde{\xi}_{+}\left(k\right) &\approx\tilde{\xi}_{+}\left(0\right)  + \frac{\hbar^{2}k^{2}}{2m^{\star}}\text{,} \\
\tilde{\Delta}_p\left(k\right) &\approx \lambda \hbar k\text{,}
\end{align*}

\noindent
where $\frac{1}{m^{\star}} = \left.\frac{\partial^{2}\tilde{\xi}_{+}\left(k\right)}{\hbar^{2}\partial k^{2}}\right|_{k=0}$ and $\lambda =  \left.\frac{\partial\tilde{\Delta}_{p}\left(k\right)}{\hbar\partial k}\right|_{k=0}$. Then the BdG Hamiltonian for $k\approx\Gamma$ becomes

\begin{equation}
\hat{\mathcal{H}}_{\text{BdG}}^{\Gamma} = \begin{pmatrix}
\frac{\hbar^{2}k^{2}}{2m^{\star}} + \tilde{\xi}_{+}\left(0\right) & \lambda \hbar k \\
\lambda \hbar k & -\left(\frac{\hbar^{2}k^{2}}{2m^{\star}} + \tilde{\xi}_{+}\left(0\right) \right)
\end{pmatrix}\text{,}
\end{equation}

\noindent
and the corresponding BdG equation reads

\begin{equation*}
\begin{pmatrix}
\frac{\hbar^{2}k^{2}}{2m^{\star}} + \tilde{\xi}_{+}\left(0\right) & \lambda \hbar k \\
\lambda \hbar k & -\left(\frac{\hbar^{2}k^{2}}{2m^{\star}} + \tilde{\xi}_{+}\left(0\right) \right)
\end{pmatrix}\begin{pmatrix}
u_{\Gamma} \\
v_{\Gamma}
\end{pmatrix} = E\begin{pmatrix}
u_{\Gamma} \\
v_{\Gamma}
\end{pmatrix}\text{.}
\end{equation*}

\noindent
Now, we interpret $k$ as the momentum operator $k\to\hat{k} = -i\partial_{z}$ and make the ansatz

\begin{equation}
\begin{pmatrix}
u_{\Gamma}\left(z\right) \\
v_{\Gamma}\left(z\right)
\end{pmatrix} = \begin{pmatrix}
u_{\Gamma} \\
v_{\Gamma}
\end{pmatrix}e^{\kappa_{\Gamma}z}\text{.}
\end{equation}

\noindent
For the  momentum $\kappa_{\Gamma}$ we need to solve the secular equation $\det\left(\hat{\mathcal{H}}_{\text{BdG}}^{\Gamma} - E\mathds{1}\right)\overset{!}{=}0$ for any energy $E$ and we obtain 

\begin{widetext}
\begin{equation}
\kappa_{\Gamma}^{2} = 2\frac{m^{\star}\tilde{\xi}_{+}\left(0\right)}{\hbar^{2}} + 2\left(\frac{m^{\star}\lambda}{\hbar}\right)^{2} \pm\sqrt{\left(2\frac{m^{\star}E}{\hbar^{2}}\right)^{2} + 4\left(\frac{m^{\star}\lambda}{\hbar}\right)^{2} \left(\left(\frac{m^{\star}\lambda}{\hbar}\right)^{2} + 2\frac{m^{\star}\tilde{\xi}_{+}\left(0\right)}{\hbar^{2}}\right)}\text{.}
\end{equation}
\end{widetext}

\noindent
For zero energy modes the equation can be simplified

\begin{equation}
\kappa_{\Gamma} = \pm\left(\frac{m^{\star}\lambda}{\hbar}\pm\sqrt{\left(\frac{m^{\star}\lambda}{\hbar}\right)^{2} + 2\frac{m^{\star}\tilde{\xi}_{+}\left(0\right)}{\hbar^{2}}}\right)\text{.}
\end{equation}

\noindent
The corresponding zero energy eigenvectors are given by

\begin{equation}
\begin{pmatrix}
u_{\Gamma} \\
v_{\Gamma}
\end{pmatrix} =\frac{1}{\sqrt{2}}\begin{pmatrix}
\mp i \\
1
\end{pmatrix}\text{.}
\end{equation}

\subsection{Fermi point contribution}

\noindent
For the Fermi point contribution we need to linearize $\tilde{\xi}_{+}\left(k\right)$ around $k_{F}$ and $-k_{F}$, see Fig. \ref{fig:LowEnergyHam}. Then, we can define the following two Nambu spinors $\Psi_{R} = \left(d_{k+,R}\text{, }d_{-k+,L}^\dag\right)$ and $\Psi_{L} = \left(d_{k+,L}\text{, }d_{-k+,R}^\dag\right)$. The subscripts $R,L$ denote the right- and left-movers. The corresponding BdG Hamiltonians are given by

\begin{equation}
\hat{\mathcal{H}}_{\text{BdG}}^{R} = \begin{pmatrix}
v_{F}\hbar\left(k - k_{F}\right) &\tilde{\Delta}_p\left(k_{F}\right) \\
\tilde{\Delta}_p\left(k_{F}\right) & -v_{F}\hbar\left(k - k_{F}\right) 
\end{pmatrix}\text{,}
\end{equation}

\begin{equation}
\hat{\mathcal{H}}_{\text{BdG}}^{L} = \begin{pmatrix}
-v_{F}\hbar\left(k + k_{F}\right) &\tilde{\Delta}_p\left(-k_{F}\right) \\
\tilde{\Delta}_p\left(-k_{F}\right) & v_{F}\hbar\left(k + k_{F}\right) 
\end{pmatrix}\text{,}
\end{equation}

\noindent
where for $\hat{\mathcal{H}}_{\text{BdG}}^{R}$ we have $k > 0$ and for $\hat{\mathcal{H}}_{\text{BdG}}^{L}$ we have $k < 0$. The corresponding BdG equation reads

\begin{equation*}
\begin{pmatrix}
v_{F}\hbar\left(k - k_{F}\right) &\tilde{\Delta}_p\left(k_{F}\right) \\
\tilde{\Delta}_p\left(k_{F}\right) & -v_{F}\hbar\left(k - k_{F}\right) 
\end{pmatrix}\begin{pmatrix}
u_{R} \\
v_{L}
\end{pmatrix} = E\begin{pmatrix}
u_{R} \\
v_{L}
\end{pmatrix}\text{,}
\end{equation*}

\begin{equation*}
\begin{pmatrix}
-v_{F}\hbar\left(k + k_{F}\right) &\tilde{\Delta}_p\left(-k_{F}\right) \\
\tilde{\Delta}_p\left(-k_{F}\right) & v_{F}\hbar\left(k + k_{F}\right) 
\end{pmatrix}\begin{pmatrix}
u_{L} \\
v_{R}
\end{pmatrix} = E\begin{pmatrix}
u_{L} \\
v_{R}
\end{pmatrix}\text{.}
\end{equation*}

\noindent
With $k\to\hat{k} = -i\partial_{z}$ and making the ansatz

\begin{equation*}
\begin{pmatrix}
u_{R}\left(z\right) \\
v_{L}\left(z\right)
\end{pmatrix} = \begin{pmatrix}
u_{R}\\
v_{L}
\end{pmatrix}e^{\kappa_{R}z}
\text{ and }\begin{pmatrix}
u_{L}\left(z\right) \\
v_{R}\left(z\right)
\end{pmatrix} = \begin{pmatrix}
u_{L}\\
v_{R}
\end{pmatrix}e^{\kappa_{L}z}\text{,}
\end{equation*}

\noindent
we get the decay lengths $\kappa_{R}$ and $\kappa_{L}$ from the secular equations $\det\left(\hat{\mathcal{H}}_{\text{BdG}}^{\text{R/L}} - E\mathds{1}\right)\overset{!}{=}0$. The decay lengths for the zero energy modes become $\kappa_{R} = ik_{F} \mp\frac{\left|\tilde{\Delta}_{p}\left(k_{F}\right)\right|}{v_{F}\hbar}$ and $\kappa_{L} = -ik_{F} \mp\frac{\left|\tilde{\Delta}_{p}\left(k_{F}\right)\right|}{v_{F}\hbar}$. Furthermore, we get the two eigenvectors

\begin{align*}
\begin{pmatrix}
u_{R} \\
v_{L}
\end{pmatrix} &= \frac{1}{\sqrt{2}}\begin{pmatrix}
\pm i\text{sgn}\left(\tilde{\Delta}_p\left(k_{F}\right)\right) \\
1
\end{pmatrix} = \frac{1}{\sqrt{2}}\begin{pmatrix}
\mp i \\
1
\end{pmatrix}\text{,} \\
\begin{pmatrix}
u_{L} \\
v_{R}
\end{pmatrix} &= \frac{1}{\sqrt{2}}\begin{pmatrix}
\mp i\text{sgn}\left(\tilde{\Delta}_p\left(-k_{F}\right)\right) \\
1
\end{pmatrix} = \frac{1}{\sqrt{2}}\begin{pmatrix}
\mp i \\
1
\end{pmatrix}\text{,}
\end{align*}
\noindent
where we used $\text{sgn}\left(\tilde{\Delta}_p\left(k_{F}\right)\right)= -1$ and $\text{sgn}\left(\tilde{\Delta}_p\left(-k_{F}\right)\right)= +1$, see Fig. \ref{fig:Setup}(b).\\

\section{Construction of 3D Majorana wave function}
\label{sec:MajoTrans}

\noindent
Explicitly, the coefficients of the electron and holes are parts of the Majorana bound state ~\eqref{eq:MajoranaValleySpin} given by

\begin{align*}
u_{K\uparrow}\left(k\right) &= a_{\uparrow}\left(k\right)\left[um\left(k\right)g\left(k\right) - vn\left(k\right)h\left(k\right)\right]\text{,} \\
u_{K^{\prime}\uparrow}\left(k\right) &= -b_{\uparrow}\left(k\right)\left[um\left(k\right)g\left(k\right) - vn\left(k\right)h\left(k\right)\right]\text{,} \\
v_{K\uparrow}\left(k\right) &= a_{\uparrow}\left(k\right)\left[vm\left(k\right)g\left(k\right) - un\left(k\right)h\left(k\right)\right]\text{,} \\
v_{K^{\prime}\uparrow}\left(k\right) &= -b_{\uparrow}\left(k\right)\left[vm\left(k\right)g\left(k\right) - un\left(k\right)h\left(k\right)\right]\text{,} \\
u_{K\downarrow}\left(k\right) &= -a_{\downarrow}\left(k\right)\left[um\left(k\right)h\left(k\right) + vn\left(k\right)g\left(k\right)\right]\text{,} \\
u_{K^{\prime}\downarrow}\left(k\right) &= b_{\downarrow}\left(k\right)\left[um\left(k\right)h\left(k\right) + vn\left(k\right)g\left(k\right)\right]\text{,} \\
v_{K\downarrow}\left(k\right) &= -a_{\downarrow}\left(k\right)\left[vm\left(k\right)h\left(k\right) + un\left(k\right)g\left(k\right)\right]\text{,} \\
v_{K^{\prime}\downarrow}\left(k\right) &= b_{\downarrow}\left(k\right)\left[vm\left(k\right)h\left(k\right) + un\left(k\right)g\left(k\right)\right]\text{.}
\end{align*}

%\bibliographystyle{apsrev4-1}
%\bibliography{transverse-bibliography}

\begin{thebibliography}{45}%
\makeatletter
\providecommand \@ifxundefined [1]{%
 \@ifx{#1\undefined}
}%
\providecommand \@ifnum [1]{%
 \ifnum #1\expandafter \@firstoftwo
 \else \expandafter \@secondoftwo
 \fi
}%
\providecommand \@ifx [1]{%
 \ifx #1\expandafter \@firstoftwo
 \else \expandafter \@secondoftwo
 \fi
}%
\providecommand \natexlab [1]{#1}%
\providecommand \enquote  [1]{``#1''}%
\providecommand \bibnamefont  [1]{#1}%
\providecommand \bibfnamefont [1]{#1}%
\providecommand \citenamefont [1]{#1}%
\providecommand \href@noop [0]{\@secondoftwo}%
\providecommand \href [0]{\begingroup \@sanitize@url \@href}%
\providecommand \@href[1]{\@@startlink{#1}\@@href}%
\providecommand \@@href[1]{\endgroup#1\@@endlink}%
\providecommand \@sanitize@url [0]{\catcode `\\12\catcode `\$12\catcode
  `\&12\catcode `\#12\catcode `\^12\catcode `\_12\catcode `\%12\relax}%
\providecommand \@@startlink[1]{}%
\providecommand \@@endlink[0]{}%
\providecommand \url  [0]{\begingroup\@sanitize@url \@url }%
\providecommand \@url [1]{\endgroup\@href {#1}{\urlprefix }}%
\providecommand \urlprefix  [0]{URL }%
\providecommand \Eprint [0]{\href }%
\providecommand \doibase [0]{http://dx.doi.org/}%
\providecommand \selectlanguage [0]{\@gobble}%
\providecommand \bibinfo  [0]{\@secondoftwo}%
\providecommand \bibfield  [0]{\@secondoftwo}%
\providecommand \translation [1]{[#1]}%
\providecommand \BibitemOpen [0]{}%
\providecommand \bibitemStop [0]{}%
\providecommand \bibitemNoStop [0]{.\EOS\space}%
\providecommand \EOS [0]{\spacefactor3000\relax}%
\providecommand \BibitemShut  [1]{\csname bibitem#1\endcsname}%
\let\auto@bib@innerbib\@empty
%</preamble>
\bibitem [{\citenamefont {Aguado}(2017)}]{Aguado-RNC-2017}%
  \BibitemOpen
  \bibfield  {author} {\bibinfo {author} {\bibfnamefont {R.}~\bibnamefont
  {Aguado}},\ }\href {\doibase 10.1393/ncr/i2017-10141-9} {\bibfield  {journal}
  {\bibinfo  {journal} {La Rivista del Nuovo Cimento}\ }\textbf {\bibinfo
  {volume} {40}},\ \bibinfo {pages} {523} (\bibinfo {year} {2017})}\BibitemShut
  {NoStop}%
\bibitem [{\citenamefont {Kitaev}(2001)}]{Kitaev-PhysUsp-2001}%
  \BibitemOpen
  \bibfield  {author} {\bibinfo {author} {\bibfnamefont {A.~Y.}\ \bibnamefont
  {Kitaev}},\ }\href {\doibase 10.1070/1063-7869/44/10S/S29} {\bibfield
  {journal} {\bibinfo  {journal} {Physics-Uspekhi}\ }\textbf {\bibinfo {volume}
  {44}},\ \bibinfo {pages} {131} (\bibinfo {year} {2001})}\BibitemShut
  {NoStop}%
\bibitem [{\citenamefont {Sato}\ and\ \citenamefont
  {Ando}(2017)}]{Sato-RPP-2017}%
  \BibitemOpen
  \bibfield  {author} {\bibinfo {author} {\bibfnamefont {M.}~\bibnamefont
  {Sato}}\ and\ \bibinfo {author} {\bibfnamefont {Y.}~\bibnamefont {Ando}},\
  }\href {\doibase 10.1088/1361-6633/aa6ac7} {\bibfield  {journal} {\bibinfo
  {journal} {Reports on Progress in Physics}\ }\textbf {\bibinfo {volume}
  {80}},\ \bibinfo {pages} {076501} (\bibinfo {year} {2017})}\BibitemShut
  {NoStop}%
\bibitem [{\citenamefont {Lutchyn}\ \emph {et~al.}(2010)\citenamefont
  {Lutchyn}, \citenamefont {Sau},\ and\ \citenamefont
  {Das~Sarma}}]{Lutchyn-PRL-2010}%
  \BibitemOpen
  \bibfield  {author} {\bibinfo {author} {\bibfnamefont {R.~M.}\ \bibnamefont
  {Lutchyn}}, \bibinfo {author} {\bibfnamefont {J.~D.}\ \bibnamefont {Sau}}, \
  and\ \bibinfo {author} {\bibfnamefont {S.}~\bibnamefont {Das~Sarma}},\ }\href
  {\doibase 10.1103/PhysRevLett.105.077001} {\bibfield  {journal} {\bibinfo
  {journal} {Phys. Rev. Lett.}\ }\textbf {\bibinfo {volume} {105}},\ \bibinfo
  {pages} {077001} (\bibinfo {year} {2010})}\BibitemShut {NoStop}%
\bibitem [{\citenamefont {Oreg}\ \emph {et~al.}(2010)\citenamefont {Oreg},
  \citenamefont {Refael},\ and\ \citenamefont {von Oppen}}]{Oreg-PRL-2010}%
  \BibitemOpen
  \bibfield  {author} {\bibinfo {author} {\bibfnamefont {Y.}~\bibnamefont
  {Oreg}}, \bibinfo {author} {\bibfnamefont {G.}~\bibnamefont {Refael}}, \ and\
  \bibinfo {author} {\bibfnamefont {F.}~\bibnamefont {von Oppen}},\ }\href
  {\doibase 10.1103/PhysRevLett.105.177002} {\bibfield  {journal} {\bibinfo
  {journal} {Phys. Rev. Lett.}\ }\textbf {\bibinfo {volume} {105}},\ \bibinfo
  {pages} {177002} (\bibinfo {year} {2010})}\BibitemShut {NoStop}%
\bibitem [{\citenamefont {Lutchyn}\ \emph {et~al.}(2018)\citenamefont
  {Lutchyn}, \citenamefont {Bakkers}, \citenamefont {Kouwenhoven},
  \citenamefont {Krogstrup}, \citenamefont {Marcus},\ and\ \citenamefont
  {Oreg}}]{Lutchyn-NatRevMat-2018}%
  \BibitemOpen
  \bibfield  {author} {\bibinfo {author} {\bibfnamefont {R.}~\bibnamefont
  {Lutchyn}}, \bibinfo {author} {\bibfnamefont {E.}~\bibnamefont {Bakkers}},
  \bibinfo {author} {\bibfnamefont {L.}~\bibnamefont {Kouwenhoven}}, \bibinfo
  {author} {\bibfnamefont {P.}~\bibnamefont {Krogstrup}}, \bibinfo {author}
  {\bibfnamefont {C.}~\bibnamefont {Marcus}}, \ and\ \bibinfo {author}
  {\bibfnamefont {Y.}~\bibnamefont {Oreg}},\ }\href {\doibase
  10.1038/s41578-018-0003-1} {\bibfield  {journal} {\bibinfo  {journal} {Nature
  Reviews Materials}\ }\textbf {\bibinfo {volume} {3}},\ \bibinfo {pages} {52}
  (\bibinfo {year} {2018})}\BibitemShut {NoStop}%
\bibitem [{\citenamefont {Mourik}\ \emph {et~al.}(2012)\citenamefont {Mourik},
  \citenamefont {Zuo}, \citenamefont {Frolov}, \citenamefont {Plissard},
  \citenamefont {Bakkers},\ and\ \citenamefont
  {Kouwenhoven}}]{Mourik-Science-2012}%
  \BibitemOpen
  \bibfield  {author} {\bibinfo {author} {\bibfnamefont {V.}~\bibnamefont
  {Mourik}}, \bibinfo {author} {\bibfnamefont {K.}~\bibnamefont {Zuo}},
  \bibinfo {author} {\bibfnamefont {S.~M.}\ \bibnamefont {Frolov}}, \bibinfo
  {author} {\bibfnamefont {S.~R.}\ \bibnamefont {Plissard}}, \bibinfo {author}
  {\bibfnamefont {E.~P. A.~M.}\ \bibnamefont {Bakkers}}, \ and\ \bibinfo
  {author} {\bibfnamefont {L.~P.}\ \bibnamefont {Kouwenhoven}},\ }\href
  {\doibase 10.1126/science.1222360} {\bibfield  {journal} {\bibinfo  {journal}
  {Science}\ }\textbf {\bibinfo {volume} {336}},\ \bibinfo {pages} {1003}
  (\bibinfo {year} {2012})}\BibitemShut {NoStop}%
\bibitem [{\citenamefont {Churchill}\ \emph {et~al.}(2013)\citenamefont
  {Churchill}, \citenamefont {Fatemi}, \citenamefont {Grove-Rasmussen},
  \citenamefont {Deng}, \citenamefont {Caroff}, \citenamefont {Xu},\ and\
  \citenamefont {Marcus}}]{Churchill-PRB-2013}%
  \BibitemOpen
  \bibfield  {author} {\bibinfo {author} {\bibfnamefont {H.~O.~H.}\
  \bibnamefont {Churchill}}, \bibinfo {author} {\bibfnamefont {V.}~\bibnamefont
  {Fatemi}}, \bibinfo {author} {\bibfnamefont {K.}~\bibnamefont
  {Grove-Rasmussen}}, \bibinfo {author} {\bibfnamefont {M.~T.}\ \bibnamefont
  {Deng}}, \bibinfo {author} {\bibfnamefont {P.}~\bibnamefont {Caroff}},
  \bibinfo {author} {\bibfnamefont {H.~Q.}\ \bibnamefont {Xu}}, \ and\ \bibinfo
  {author} {\bibfnamefont {C.~M.}\ \bibnamefont {Marcus}},\ }\href {\doibase
  10.1103/PhysRevB.87.241401} {\bibfield  {journal} {\bibinfo  {journal} {Phys.
  Rev. B}\ }\textbf {\bibinfo {volume} {87}},\ \bibinfo {pages} {241401}
  (\bibinfo {year} {2013})}\BibitemShut {NoStop}%
\bibitem [{\citenamefont {Deng}\ \emph {et~al.}(2016)\citenamefont {Deng},
  \citenamefont {Vaitiekenas}, \citenamefont {Hansen}, \citenamefont {Danon},
  \citenamefont {Leijnse}, \citenamefont {Flensberg}, \citenamefont
  {Nyg{\aa}rd}, \citenamefont {Krogstrup},\ and\ \citenamefont
  {Marcus}}]{Deng-Science-2016}%
  \BibitemOpen
  \bibfield  {author} {\bibinfo {author} {\bibfnamefont {M.~T.}\ \bibnamefont
  {Deng}}, \bibinfo {author} {\bibfnamefont {S.}~\bibnamefont {Vaitiekenas}},
  \bibinfo {author} {\bibfnamefont {E.~B.}\ \bibnamefont {Hansen}}, \bibinfo
  {author} {\bibfnamefont {J.}~\bibnamefont {Danon}}, \bibinfo {author}
  {\bibfnamefont {M.}~\bibnamefont {Leijnse}}, \bibinfo {author} {\bibfnamefont
  {K.}~\bibnamefont {Flensberg}}, \bibinfo {author} {\bibfnamefont
  {J.}~\bibnamefont {Nyg{\aa}rd}}, \bibinfo {author} {\bibfnamefont
  {P.}~\bibnamefont {Krogstrup}}, \ and\ \bibinfo {author} {\bibfnamefont
  {C.~M.}\ \bibnamefont {Marcus}},\ }\href {\doibase 10.1126/science.aaf3961}
  {\bibfield  {journal} {\bibinfo  {journal} {Science}\ }\textbf {\bibinfo
  {volume} {354}},\ \bibinfo {pages} {1557} (\bibinfo {year}
  {2016})}\BibitemShut {NoStop}%
\bibitem [{\citenamefont {Zhang}\ \emph {et~al.}(2017)\citenamefont {Zhang},
  \citenamefont {\"{O}nder G\"{u}l}, \citenamefont {Conesa-Boj}, \citenamefont
  {Nowak}, \citenamefont {Wimmer}, \citenamefont {Zuo}, \citenamefont {Mourik},
  \citenamefont {de~Vries}, \citenamefont {van Veen}, \citenamefont {de~Moor},
  \citenamefont {Bommer}, \citenamefont {van Woerkom}, \citenamefont {Car},
  \citenamefont {Plissard}, \citenamefont {Bakkers}, \citenamefont
  {Quintero-P\'{e}rez}, \citenamefont {Cassidy}, \citenamefont {Koelling},
  \citenamefont {Goswami}, \citenamefont {Watanabe}, \citenamefont
  {Taniguchi},\ and\ \citenamefont {Kouwenhoven}}]{Zhang-NatComm-2017}%
  \BibitemOpen
  \bibfield  {author} {\bibinfo {author} {\bibfnamefont {H.}~\bibnamefont
  {Zhang}}, \bibinfo {author} {\bibnamefont {\"{O}nder G\"{u}l}}, \bibinfo
  {author} {\bibfnamefont {S.}~\bibnamefont {Conesa-Boj}}, \bibinfo {author}
  {\bibfnamefont {M.~P.}\ \bibnamefont {Nowak}}, \bibinfo {author}
  {\bibfnamefont {M.}~\bibnamefont {Wimmer}}, \bibinfo {author} {\bibfnamefont
  {K.}~\bibnamefont {Zuo}}, \bibinfo {author} {\bibfnamefont {V.}~\bibnamefont
  {Mourik}}, \bibinfo {author} {\bibfnamefont {F.~K.}\ \bibnamefont
  {de~Vries}}, \bibinfo {author} {\bibfnamefont {J.}~\bibnamefont {van Veen}},
  \bibinfo {author} {\bibfnamefont {M.~W.~A.}\ \bibnamefont {de~Moor}},
  \bibinfo {author} {\bibfnamefont {J.~D.~S.}\ \bibnamefont {Bommer}}, \bibinfo
  {author} {\bibfnamefont {D.~J.}\ \bibnamefont {van Woerkom}}, \bibinfo
  {author} {\bibfnamefont {D.}~\bibnamefont {Car}}, \bibinfo {author}
  {\bibfnamefont {S.~R.}\ \bibnamefont {Plissard}}, \bibinfo {author}
  {\bibfnamefont {E.~P.}\ \bibnamefont {Bakkers}}, \bibinfo {author}
  {\bibfnamefont {M.}~\bibnamefont {Quintero-P\'{e}rez}}, \bibinfo {author}
  {\bibfnamefont {M.~C.}\ \bibnamefont {Cassidy}}, \bibinfo {author}
  {\bibfnamefont {S.}~\bibnamefont {Koelling}}, \bibinfo {author}
  {\bibfnamefont {S.}~\bibnamefont {Goswami}}, \bibinfo {author} {\bibfnamefont
  {K.}~\bibnamefont {Watanabe}}, \bibinfo {author} {\bibfnamefont
  {T.}~\bibnamefont {Taniguchi}}, \ and\ \bibinfo {author} {\bibfnamefont
  {L.~P.}\ \bibnamefont {Kouwenhoven}},\ }\href {\doibase 10.1038/ncomms16025}
  {\bibfield  {journal} {\bibinfo  {journal} {Nature Communications}\ }\textbf
  {\bibinfo {volume} {8}},\ \bibinfo {pages} {1} (\bibinfo {year} {2017})},\
  \Eprint {http://arxiv.org/abs/http://www.nature.com/articles/ncomms16025.pdf}
  {http://www.nature.com/articles/ncomms16025.pdf} \BibitemShut {NoStop}%
\bibitem [{\citenamefont {Liu}\ \emph {et~al.}(2017)\citenamefont {Liu},
  \citenamefont {Sau}, \citenamefont {Stanescu},\ and\ \citenamefont
  {Das~Sarma}}]{Liu-PRB-2017}%
  \BibitemOpen
  \bibfield  {author} {\bibinfo {author} {\bibfnamefont {C.-X.}\ \bibnamefont
  {Liu}}, \bibinfo {author} {\bibfnamefont {J.~D.}\ \bibnamefont {Sau}},
  \bibinfo {author} {\bibfnamefont {T.~D.}\ \bibnamefont {Stanescu}}, \ and\
  \bibinfo {author} {\bibfnamefont {S.}~\bibnamefont {Das~Sarma}},\ }\href
  {\doibase 10.1103/PhysRevB.96.075161} {\bibfield  {journal} {\bibinfo
  {journal} {Phys. Rev. B}\ }\textbf {\bibinfo {volume} {96}},\ \bibinfo
  {pages} {075161} (\bibinfo {year} {2017})}\BibitemShut {NoStop}%
\bibitem [{\citenamefont {Prada}\ \emph {et~al.}(2017)\citenamefont {Prada},
  \citenamefont {Aguado},\ and\ \citenamefont {San-Jose}}]{Prada-PRB-2017}%
  \BibitemOpen
  \bibfield  {author} {\bibinfo {author} {\bibfnamefont {E.}~\bibnamefont
  {Prada}}, \bibinfo {author} {\bibfnamefont {R.}~\bibnamefont {Aguado}}, \
  and\ \bibinfo {author} {\bibfnamefont {P.}~\bibnamefont {San-Jose}},\ }\href
  {\doibase 10.1103/PhysRevB.96.085418} {\bibfield  {journal} {\bibinfo
  {journal} {Phys. Rev. B}\ }\textbf {\bibinfo {volume} {96}},\ \bibinfo
  {pages} {085418} (\bibinfo {year} {2017})}\BibitemShut {NoStop}%
\bibitem [{\citenamefont {Clarke}(2017)}]{Clarke-PRB-2017}%
  \BibitemOpen
  \bibfield  {author} {\bibinfo {author} {\bibfnamefont {D.~J.}\ \bibnamefont
  {Clarke}},\ }\href {\doibase 10.1103/PhysRevB.96.201109} {\bibfield
  {journal} {\bibinfo  {journal} {Phys. Rev. B}\ }\textbf {\bibinfo {volume}
  {96}},\ \bibinfo {pages} {201109} (\bibinfo {year} {2017})}\BibitemShut
  {NoStop}%
\bibitem [{\citenamefont {Spanton}\ \emph {et~al.}(2017)\citenamefont
  {Spanton}, \citenamefont {Deng}, \citenamefont {Vaitiek\'{e}nas},
  \citenamefont {Krogstrup}, \citenamefont {Nyg\o{a}rd}, \citenamefont
  {Marcus},\ and\ \citenamefont {Moler}}]{Deng-NatPhys-2017}%
  \BibitemOpen
  \bibfield  {author} {\bibinfo {author} {\bibfnamefont {E.~M.}\ \bibnamefont
  {Spanton}}, \bibinfo {author} {\bibfnamefont {M.}~\bibnamefont {Deng}},
  \bibinfo {author} {\bibfnamefont {S.}~\bibnamefont {Vaitiek\'{e}nas}},
  \bibinfo {author} {\bibfnamefont {P.}~\bibnamefont {Krogstrup}}, \bibinfo
  {author} {\bibfnamefont {J.}~\bibnamefont {Nyg\o{a}rd}}, \bibinfo {author}
  {\bibfnamefont {C.~M.}\ \bibnamefont {Marcus}}, \ and\ \bibinfo {author}
  {\bibfnamefont {K.~A.}\ \bibnamefont {Moler}},\ }\href {\doibase
  10.1038/nphys4224} {\bibfield  {journal} {\bibinfo  {journal} {Nature
  Physics}\ ,\ \bibinfo {pages} {1177}} (\bibinfo {year} {2017})}\BibitemShut
  {NoStop}%
\bibitem [{\citenamefont {Hoffman}\ \emph {et~al.}(2017)\citenamefont
  {Hoffman}, \citenamefont {Chevallier}, \citenamefont {Loss},\ and\
  \citenamefont {Klinovaja}}]{Hoffman-PRB-2017}%
  \BibitemOpen
  \bibfield  {author} {\bibinfo {author} {\bibfnamefont {S.}~\bibnamefont
  {Hoffman}}, \bibinfo {author} {\bibfnamefont {D.}~\bibnamefont {Chevallier}},
  \bibinfo {author} {\bibfnamefont {D.}~\bibnamefont {Loss}}, \ and\ \bibinfo
  {author} {\bibfnamefont {J.}~\bibnamefont {Klinovaja}},\ }\href {\doibase
  10.1103/PhysRevB.96.045440} {\bibfield  {journal} {\bibinfo  {journal} {Phys.
  Rev. B}\ }\textbf {\bibinfo {volume} {96}},\ \bibinfo {pages} {045440}
  (\bibinfo {year} {2017})}\BibitemShut {NoStop}%
\bibitem [{\citenamefont {Schuray}\ \emph {et~al.}(2018)\citenamefont
  {Schuray}, \citenamefont {Yeyati},\ and\ \citenamefont
  {Recher}}]{Schuray-PRB-2018}%
  \BibitemOpen
  \bibfield  {author} {\bibinfo {author} {\bibfnamefont {A.}~\bibnamefont
  {Schuray}}, \bibinfo {author} {\bibfnamefont {A.~L.}\ \bibnamefont {Yeyati}},
  \ and\ \bibinfo {author} {\bibfnamefont {P.}~\bibnamefont {Recher}},\ }\href
  {\doibase 10.1103/PhysRevB.98.235301} {\bibfield  {journal} {\bibinfo
  {journal} {Phys. Rev. B}\ }\textbf {\bibinfo {volume} {98}},\ \bibinfo
  {pages} {235301} (\bibinfo {year} {2018})}\BibitemShut {NoStop}%
\bibitem [{\citenamefont {Sticlet}\ \emph {et~al.}(2012)\citenamefont
  {Sticlet}, \citenamefont {Bena},\ and\ \citenamefont
  {Simon}}]{Bena-PRL-2012}%
  \BibitemOpen
  \bibfield  {author} {\bibinfo {author} {\bibfnamefont {D.}~\bibnamefont
  {Sticlet}}, \bibinfo {author} {\bibfnamefont {C.}~\bibnamefont {Bena}}, \
  and\ \bibinfo {author} {\bibfnamefont {P.}~\bibnamefont {Simon}},\ }\href
  {\doibase 10.1103/PhysRevLett.108.096802} {\bibfield  {journal} {\bibinfo
  {journal} {Phys. Rev. Lett.}\ }\textbf {\bibinfo {volume} {108}},\ \bibinfo
  {pages} {096802} (\bibinfo {year} {2012})}\BibitemShut {NoStop}%
\bibitem [{\citenamefont {Sedlmayr}\ and\ \citenamefont
  {Bena}(2015)}]{Bena-PRB-2015}%
  \BibitemOpen
  \bibfield  {author} {\bibinfo {author} {\bibfnamefont {N.}~\bibnamefont
  {Sedlmayr}}\ and\ \bibinfo {author} {\bibfnamefont {C.}~\bibnamefont
  {Bena}},\ }\href {\doibase 10.1103/PhysRevB.92.115115} {\bibfield  {journal}
  {\bibinfo  {journal} {Phys. Rev. B}\ }\textbf {\bibinfo {volume} {92}},\
  \bibinfo {pages} {115115} (\bibinfo {year} {2015})}\BibitemShut {NoStop}%
\bibitem [{\citenamefont {Klinovaja}\ and\ \citenamefont
  {Loss}(2012)}]{Klinovaja-PRB-2012}%
  \BibitemOpen
  \bibfield  {author} {\bibinfo {author} {\bibfnamefont {J.}~\bibnamefont
  {Klinovaja}}\ and\ \bibinfo {author} {\bibfnamefont {D.}~\bibnamefont
  {Loss}},\ }\href {\doibase 10.1103/PhysRevB.86.085408} {\bibfield  {journal}
  {\bibinfo  {journal} {Phys. Rev. B}\ }\textbf {\bibinfo {volume} {86}},\
  \bibinfo {pages} {085408} (\bibinfo {year} {2012})}\BibitemShut {NoStop}%
\bibitem [{\citenamefont {Lim}\ \emph {et~al.}(2013)\citenamefont {Lim},
  \citenamefont {Lopez},\ and\ \citenamefont {Serra}}]{Lim-EPL-2013}%
  \BibitemOpen
  \bibfield  {author} {\bibinfo {author} {\bibfnamefont {J.~S.}\ \bibnamefont
  {Lim}}, \bibinfo {author} {\bibfnamefont {R.}~\bibnamefont {Lopez}}, \ and\
  \bibinfo {author} {\bibfnamefont {L.}~\bibnamefont {Serra}},\ }\href
  {http://stacks.iop.org/0295-5075/103/i=3/a=37004} {\bibfield  {journal}
  {\bibinfo  {journal} {EPL (Europhysics Letters)}\ }\textbf {\bibinfo {volume}
  {103}},\ \bibinfo {pages} {37004} (\bibinfo {year} {2013})}\BibitemShut
  {NoStop}%
\bibitem [{\citenamefont {Osca}\ \emph {et~al.}(2014)\citenamefont {Osca},
  \citenamefont {L{\'o}pez},\ and\ \citenamefont {Serra}}]{Osca-EPJB-2014}%
  \BibitemOpen
  \bibfield  {author} {\bibinfo {author} {\bibfnamefont {J.}~\bibnamefont
  {Osca}}, \bibinfo {author} {\bibfnamefont {R.}~\bibnamefont {L{\'o}pez}}, \
  and\ \bibinfo {author} {\bibfnamefont {L.}~\bibnamefont {Serra}},\ }\href
  {\doibase 10.1140/epjb/e2014-41091-8} {\bibfield  {journal} {\bibinfo
  {journal} {The European Physical Journal B}\ }\textbf {\bibinfo {volume}
  {87}},\ \bibinfo {pages} {84} (\bibinfo {year} {2014})}\BibitemShut {NoStop}%
\bibitem [{\citenamefont {Manolescu}\ \emph {et~al.}(2017)\citenamefont
  {Manolescu}, \citenamefont {Sitek}, \citenamefont {Osca}, \citenamefont
  {Serra}, \citenamefont {Gudmundsson},\ and\ \citenamefont
  {Stanescu}}]{Manolescu-PRB-2017}%
  \BibitemOpen
  \bibfield  {author} {\bibinfo {author} {\bibfnamefont {A.}~\bibnamefont
  {Manolescu}}, \bibinfo {author} {\bibfnamefont {A.}~\bibnamefont {Sitek}},
  \bibinfo {author} {\bibfnamefont {J.}~\bibnamefont {Osca}}, \bibinfo {author}
  {\bibfnamefont {L.}~\bibnamefont {Serra}}, \bibinfo {author} {\bibfnamefont
  {V.}~\bibnamefont {Gudmundsson}}, \ and\ \bibinfo {author} {\bibfnamefont
  {T.~D.}\ \bibnamefont {Stanescu}},\ }\href {\doibase
  10.1103/PhysRevB.96.125435} {\bibfield  {journal} {\bibinfo  {journal} {Phys.
  Rev. B}\ }\textbf {\bibinfo {volume} {96}},\ \bibinfo {pages} {125435}
  (\bibinfo {year} {2017})}\BibitemShut {NoStop}%
\bibitem [{\citenamefont {Stanescu}\ \emph {et~al.}(2018)\citenamefont
  {Stanescu}, \citenamefont {Sitek},\ and\ \citenamefont
  {Manolescu}}]{Stanescu-BJN-2018}%
  \BibitemOpen
  \bibfield  {author} {\bibinfo {author} {\bibfnamefont {T.~D.}\ \bibnamefont
  {Stanescu}}, \bibinfo {author} {\bibfnamefont {A.}~\bibnamefont {Sitek}}, \
  and\ \bibinfo {author} {\bibfnamefont {A.}~\bibnamefont {Manolescu}},\ }\href
  {\doibase 10.3762/bjnano.9.142} {\bibfield  {journal} {\bibinfo  {journal}
  {Beilstein J. Nanotechnol.}\ }\textbf {\bibinfo {volume} {9}},\ \bibinfo
  {pages} {1512} (\bibinfo {year} {2018})}\BibitemShut {NoStop}%
\bibitem [{\citenamefont {Woods}\ \emph {et~al.}(2018)\citenamefont {Woods},
  \citenamefont {Stanescu},\ and\ \citenamefont {Das~Sarma}}]{Woods-PRB-2018}%
  \BibitemOpen
  \bibfield  {author} {\bibinfo {author} {\bibfnamefont {B.~D.}\ \bibnamefont
  {Woods}}, \bibinfo {author} {\bibfnamefont {T.~D.}\ \bibnamefont {Stanescu}},
  \ and\ \bibinfo {author} {\bibfnamefont {S.}~\bibnamefont {Das~Sarma}},\
  }\href {\doibase 10.1103/PhysRevB.98.035428} {\bibfield  {journal} {\bibinfo
  {journal} {Phys. Rev. B}\ }\textbf {\bibinfo {volume} {98}},\ \bibinfo
  {pages} {035428} (\bibinfo {year} {2018})}\BibitemShut {NoStop}%
\bibitem [{\citenamefont {Egger}\ and\ \citenamefont
  {Flensberg}(2012)}]{Egger-PRB-2012}%
  \BibitemOpen
  \bibfield  {author} {\bibinfo {author} {\bibfnamefont {R.}~\bibnamefont
  {Egger}}\ and\ \bibinfo {author} {\bibfnamefont {K.}~\bibnamefont
  {Flensberg}},\ }\href {\doibase 10.1103/PhysRevB.85.235462} {\bibfield
  {journal} {\bibinfo  {journal} {Phys. Rev. B}\ }\textbf {\bibinfo {volume}
  {85}},\ \bibinfo {pages} {235462} (\bibinfo {year} {2012})}\BibitemShut
  {NoStop}%
\bibitem [{\citenamefont {Klinovaja}\ \emph {et~al.}(2012)\citenamefont
  {Klinovaja}, \citenamefont {Gangadharaiah},\ and\ \citenamefont
  {Loss}}]{Klinovaja-PRL-2012}%
  \BibitemOpen
  \bibfield  {author} {\bibinfo {author} {\bibfnamefont {J.}~\bibnamefont
  {Klinovaja}}, \bibinfo {author} {\bibfnamefont {S.}~\bibnamefont
  {Gangadharaiah}}, \ and\ \bibinfo {author} {\bibfnamefont {D.}~\bibnamefont
  {Loss}},\ }\href {\doibase 10.1103/PhysRevLett.108.196804} {\bibfield
  {journal} {\bibinfo  {journal} {Phys. Rev. Lett.}\ }\textbf {\bibinfo
  {volume} {108}},\ \bibinfo {pages} {196804} (\bibinfo {year}
  {2012})}\BibitemShut {NoStop}%
\bibitem [{\citenamefont {Sau}\ and\ \citenamefont
  {Tewari}(2013)}]{Sau-PRB-2013}%
  \BibitemOpen
  \bibfield  {author} {\bibinfo {author} {\bibfnamefont {J.~D.}\ \bibnamefont
  {Sau}}\ and\ \bibinfo {author} {\bibfnamefont {S.}~\bibnamefont {Tewari}},\
  }\href {\doibase 10.1103/PhysRevB.88.054503} {\bibfield  {journal} {\bibinfo
  {journal} {Phys. Rev. B}\ }\textbf {\bibinfo {volume} {88}},\ \bibinfo
  {pages} {054503} (\bibinfo {year} {2013})}\BibitemShut {NoStop}%
\bibitem [{\citenamefont {Hsu}\ \emph {et~al.}(2015)\citenamefont {Hsu},
  \citenamefont {Stano}, \citenamefont {Klinovaja},\ and\ \citenamefont
  {Loss}}]{Hsu-PRB-2015}%
  \BibitemOpen
  \bibfield  {author} {\bibinfo {author} {\bibfnamefont {C.-H.}\ \bibnamefont
  {Hsu}}, \bibinfo {author} {\bibfnamefont {P.}~\bibnamefont {Stano}}, \bibinfo
  {author} {\bibfnamefont {J.}~\bibnamefont {Klinovaja}}, \ and\ \bibinfo
  {author} {\bibfnamefont {D.}~\bibnamefont {Loss}},\ }\href {\doibase
  10.1103/PhysRevB.92.235435} {\bibfield  {journal} {\bibinfo  {journal} {Phys.
  Rev. B}\ }\textbf {\bibinfo {volume} {92}},\ \bibinfo {pages} {235435}
  (\bibinfo {year} {2015})}\BibitemShut {NoStop}%
\bibitem [{\citenamefont {Marganska}\ \emph {et~al.}(2018)\citenamefont
  {Marganska}, \citenamefont {Milz}, \citenamefont {Izumida}, \citenamefont
  {Strunk},\ and\ \citenamefont {Grifoni}}]{Milz-PRB-2018}%
  \BibitemOpen
  \bibfield  {author} {\bibinfo {author} {\bibfnamefont {M.}~\bibnamefont
  {Marganska}}, \bibinfo {author} {\bibfnamefont {L.}~\bibnamefont {Milz}},
  \bibinfo {author} {\bibfnamefont {W.}~\bibnamefont {Izumida}}, \bibinfo
  {author} {\bibfnamefont {C.}~\bibnamefont {Strunk}}, \ and\ \bibinfo {author}
  {\bibfnamefont {M.}~\bibnamefont {Grifoni}},\ }\href {\doibase
  10.1103/PhysRevB.97.075141} {\bibfield  {journal} {\bibinfo  {journal} {Phys.
  Rev. B}\ }\textbf {\bibinfo {volume} {97}},\ \bibinfo {pages} {075141}
  (\bibinfo {year} {2018})}\BibitemShut {NoStop}%
\bibitem [{\citenamefont {Izumida}\ \emph {et~al.}(2009)\citenamefont
  {Izumida}, \citenamefont {Sato},\ and\ \citenamefont
  {Saito}}]{Izumida-JPSJ-2009}%
  \BibitemOpen
  \bibfield  {author} {\bibinfo {author} {\bibfnamefont {W.}~\bibnamefont
  {Izumida}}, \bibinfo {author} {\bibfnamefont {K.}~\bibnamefont {Sato}}, \
  and\ \bibinfo {author} {\bibfnamefont {R.}~\bibnamefont {Saito}},\ }\href
  {\doibase 10.1143/JPSJ.78.074707} {\bibfield  {journal} {\bibinfo  {journal}
  {Journal of the Physical Society of Japan}\ }\textbf {\bibinfo {volume}
  {78}},\ \bibinfo {pages} {074707} (\bibinfo {year} {2009})}\BibitemShut
  {NoStop}%
\bibitem [{\citenamefont {Klinovaja}\ \emph {et~al.}(2011)\citenamefont
  {Klinovaja}, \citenamefont {Schmidt}, \citenamefont {Braunecker},\ and\
  \citenamefont {Loss}}]{Klinovaja-PRB-2011}%
  \BibitemOpen
  \bibfield  {author} {\bibinfo {author} {\bibfnamefont {J.}~\bibnamefont
  {Klinovaja}}, \bibinfo {author} {\bibfnamefont {M.~J.}\ \bibnamefont
  {Schmidt}}, \bibinfo {author} {\bibfnamefont {B.}~\bibnamefont {Braunecker}},
  \ and\ \bibinfo {author} {\bibfnamefont {D.}~\bibnamefont {Loss}},\ }\href
  {\doibase 10.1103/PhysRevB.84.085452} {\bibfield  {journal} {\bibinfo
  {journal} {Phys. Rev. B}\ }\textbf {\bibinfo {volume} {84}},\ \bibinfo
  {pages} {085452} (\bibinfo {year} {2011})}\BibitemShut {NoStop}%
\bibitem [{\citenamefont {Ando}(2000)}]{Ando-JPSJ-2000}%
  \BibitemOpen
  \bibfield  {author} {\bibinfo {author} {\bibfnamefont {T.}~\bibnamefont
  {Ando}},\ }\href {\doibase 10.1143/JPSJ.69.1757} {\bibfield  {journal}
  {\bibinfo  {journal} {Journal of the Physical Society of Japan}\ }\textbf
  {\bibinfo {volume} {69}},\ \bibinfo {pages} {1757} (\bibinfo {year}
  {2000})}\BibitemShut {NoStop}%
\bibitem [{Note1()}]{Note1}%
  \BibitemOpen
  \bibinfo {note} {In our tight binding model we consider one $p_{z}$ orbital
  per atom.}\BibitemShut {Stop}%
\bibitem [{\citenamefont {Saito}\ \emph {et~al.}(1998)\citenamefont {Saito},
  \citenamefont {Dresselhaus},\ and\ \citenamefont {Dresselhaus}}]{SaitoBook}%
  \BibitemOpen
  \bibfield  {author} {\bibinfo {author} {\bibfnamefont {R.}~\bibnamefont
  {Saito}}, \bibinfo {author} {\bibfnamefont {G.}~\bibnamefont {Dresselhaus}},
  \ and\ \bibinfo {author} {\bibfnamefont {M.~S.}\ \bibnamefont
  {Dresselhaus}},\ }\href@noop {} {\emph {\bibinfo {title} {Physical Properties
  of Carbon Nanotubes}}}\ (\bibinfo  {publisher} {Imperial College Press,
  London},\ \bibinfo {year} {1998})\BibitemShut {NoStop}%
\bibitem [{\citenamefont {Kuemmeth}\ \emph {et~al.}(2008)\citenamefont
  {Kuemmeth}, \citenamefont {Ilani}, \citenamefont {Ralph},\ and\ \citenamefont
  {McEuen}}]{Kuemmeth-Nature-2008}%
  \BibitemOpen
  \bibfield  {author} {\bibinfo {author} {\bibfnamefont {F.}~\bibnamefont
  {Kuemmeth}}, \bibinfo {author} {\bibfnamefont {S.}~\bibnamefont {Ilani}},
  \bibinfo {author} {\bibfnamefont {D.}~\bibnamefont {Ralph}}, \ and\ \bibinfo
  {author} {\bibfnamefont {P.}~\bibnamefont {McEuen}},\ }\href {\doibase
  10.1038/nature06822} {\bibfield  {journal} {\bibinfo  {journal} {Nature}\
  }\textbf {\bibinfo {volume} {452}},\ \bibinfo {pages} {448} (\bibinfo {year}
  {2008})}\BibitemShut {NoStop}%
\bibitem [{\citenamefont {Marganska}\ \emph {et~al.}(2015)\citenamefont
  {Marganska}, \citenamefont {Chudzinski},\ and\ \citenamefont
  {Grifoni}}]{Marganska-PRB-2015}%
  \BibitemOpen
  \bibfield  {author} {\bibinfo {author} {\bibfnamefont {M.}~\bibnamefont
  {Marganska}}, \bibinfo {author} {\bibfnamefont {P.}~\bibnamefont
  {Chudzinski}}, \ and\ \bibinfo {author} {\bibfnamefont {M.}~\bibnamefont
  {Grifoni}},\ }\href {\doibase 10.1103/PhysRevB.92.075433} {\bibfield
  {journal} {\bibinfo  {journal} {Phys. Rev. B}\ }\textbf {\bibinfo {volume}
  {92}},\ \bibinfo {pages} {075433} (\bibinfo {year} {2015})}\BibitemShut
  {NoStop}%
\bibitem [{\citenamefont {Izumida}\ \emph {et~al.}(2016)\citenamefont
  {Izumida}, \citenamefont {Okuyama}, \citenamefont {Yamakage},\ and\
  \citenamefont {Saito}}]{Izumida-PRB-2016}%
  \BibitemOpen
  \bibfield  {author} {\bibinfo {author} {\bibfnamefont {W.}~\bibnamefont
  {Izumida}}, \bibinfo {author} {\bibfnamefont {R.}~\bibnamefont {Okuyama}},
  \bibinfo {author} {\bibfnamefont {A.}~\bibnamefont {Yamakage}}, \ and\
  \bibinfo {author} {\bibfnamefont {R.}~\bibnamefont {Saito}},\ }\href
  {\doibase 10.1103/PhysRevB.93.195442} {\bibfield  {journal} {\bibinfo
  {journal} {Phys. Rev. B}\ }\textbf {\bibinfo {volume} {93}},\ \bibinfo
  {pages} {195442} (\bibinfo {year} {2016})}\BibitemShut {NoStop}%
\bibitem [{\citenamefont {Uchoa}\ and\ \citenamefont
  {Castro~Neto}(2007)}]{Uchoa-PRL-2007}%
  \BibitemOpen
  \bibfield  {author} {\bibinfo {author} {\bibfnamefont {B.}~\bibnamefont
  {Uchoa}}\ and\ \bibinfo {author} {\bibfnamefont {A.~H.}\ \bibnamefont
  {Castro~Neto}},\ }\href {\doibase 10.1103/PhysRevLett.98.146801} {\bibfield
  {journal} {\bibinfo  {journal} {Phys. Rev. Lett.}\ }\textbf {\bibinfo
  {volume} {98}},\ \bibinfo {pages} {146801} (\bibinfo {year}
  {2007})}\BibitemShut {NoStop}%
\bibitem [{\citenamefont {Shiozaki}\ and\ \citenamefont
  {Sato}(2014)}]{Shiozaki-PRB-2014}%
  \BibitemOpen
  \bibfield  {author} {\bibinfo {author} {\bibfnamefont {K.}~\bibnamefont
  {Shiozaki}}\ and\ \bibinfo {author} {\bibfnamefont {M.}~\bibnamefont
  {Sato}},\ }\href {\doibase 10.1103/PhysRevB.90.165114} {\bibfield  {journal}
  {\bibinfo  {journal} {Phys. Rev. B}\ }\textbf {\bibinfo {volume} {90}},\
  \bibinfo {pages} {165114} (\bibinfo {year} {2014})}\BibitemShut {NoStop}%
\bibitem [{\citenamefont {Ando}\ and\ \citenamefont
  {Fu}(2015)}]{Ando-AnnuRev-2015}%
  \BibitemOpen
  \bibfield  {author} {\bibinfo {author} {\bibfnamefont {Y.}~\bibnamefont
  {Ando}}\ and\ \bibinfo {author} {\bibfnamefont {L.}~\bibnamefont {Fu}},\
  }\href {\doibase 10.1146/annurev-conmatphys-031214-014501} {\bibfield
  {journal} {\bibinfo  {journal} {Annual Review of Condensed Matter Physics}\
  }\textbf {\bibinfo {volume} {6}},\ \bibinfo {pages} {361} (\bibinfo {year}
  {2015})}\BibitemShut {NoStop}%
\bibitem [{\citenamefont {del Valle}\ \emph {et~al.}(2011)\citenamefont {del
  Valle}, \citenamefont {Marga\'{n}ska},\ and\ \citenamefont
  {Grifoni}}]{delValle-PRB-2011}%
  \BibitemOpen
  \bibfield  {author} {\bibinfo {author} {\bibfnamefont {M.}~\bibnamefont {del
  Valle}}, \bibinfo {author} {\bibfnamefont {M.}~\bibnamefont {Marga\'{n}ska}},
  \ and\ \bibinfo {author} {\bibfnamefont {M.}~\bibnamefont {Grifoni}},\ }\href
  {\doibase 10.1103/PhysRevB.84.165427} {\bibfield  {journal} {\bibinfo
  {journal} {Phys. Rev. B}\ }\textbf {\bibinfo {volume} {84}},\ \bibinfo
  {pages} {165427} (\bibinfo {year} {2011})}\BibitemShut {NoStop}%
\bibitem [{\citenamefont {Cao}\ \emph {et~al.}(2005)\citenamefont {Cao},
  \citenamefont {Wang},\ and\ \citenamefont {Dai}}]{Cao-NatMat-2005}%
  \BibitemOpen
  \bibfield  {author} {\bibinfo {author} {\bibfnamefont {J.}~\bibnamefont
  {Cao}}, \bibinfo {author} {\bibfnamefont {Q.}~\bibnamefont {Wang}}, \ and\
  \bibinfo {author} {\bibfnamefont {H.}~\bibnamefont {Dai}},\ }\href {\doibase
  10.1038/nmat1478} {\bibfield  {journal} {\bibinfo  {journal} {Nature
  Materials}\ }\textbf {\bibinfo {volume} {4}},\ \bibinfo {pages} {745}
  (\bibinfo {year} {2005})}\BibitemShut {NoStop}%
\bibitem [{\citenamefont {Deshpande}\ \emph {et~al.}(2009)\citenamefont
  {Deshpande}, \citenamefont {Chandra}, \citenamefont {Caldwell}, \citenamefont
  {Novikov}, \citenamefont {Hone},\ and\ \citenamefont
  {Bockrath}}]{Deshpande-Science-2009}%
  \BibitemOpen
  \bibfield  {author} {\bibinfo {author} {\bibfnamefont {V.~V.}\ \bibnamefont
  {Deshpande}}, \bibinfo {author} {\bibfnamefont {B.}~\bibnamefont {Chandra}},
  \bibinfo {author} {\bibfnamefont {R.}~\bibnamefont {Caldwell}}, \bibinfo
  {author} {\bibfnamefont {D.~S.}\ \bibnamefont {Novikov}}, \bibinfo {author}
  {\bibfnamefont {J.}~\bibnamefont {Hone}}, \ and\ \bibinfo {author}
  {\bibfnamefont {M.}~\bibnamefont {Bockrath}},\ }\href {\doibase
  10.1126/science.1165799} {\bibfield  {journal} {\bibinfo  {journal}
  {Science}\ }\textbf {\bibinfo {volume} {323}},\ \bibinfo {pages} {106}
  (\bibinfo {year} {2009})}\BibitemShut {NoStop}%
\bibitem [{\citenamefont {Jung}\ \emph {et~al.}(2013)\citenamefont {Jung},
  \citenamefont {Schindele}, \citenamefont {Nau}, \citenamefont {Weiss},
  \citenamefont {Baumgartner},\ and\ \citenamefont
  {Sch\"{o}nenberger}}]{Jung-Nanolett-2013}%
  \BibitemOpen
  \bibfield  {author} {\bibinfo {author} {\bibfnamefont {M.}~\bibnamefont
  {Jung}}, \bibinfo {author} {\bibfnamefont {J.}~\bibnamefont {Schindele}},
  \bibinfo {author} {\bibfnamefont {S.}~\bibnamefont {Nau}}, \bibinfo {author}
  {\bibfnamefont {M.}~\bibnamefont {Weiss}}, \bibinfo {author} {\bibfnamefont
  {A.}~\bibnamefont {Baumgartner}}, \ and\ \bibinfo {author} {\bibfnamefont
  {C.}~\bibnamefont {Sch\"{o}nenberger}},\ }\href {\doibase 10.1021/nl402455n}
  {\bibfield  {journal} {\bibinfo  {journal} {Nano Letters}\ }\textbf {\bibinfo
  {volume} {13}},\ \bibinfo {pages} {4522} (\bibinfo {year} {2013})},\ \Eprint
  {http://arxiv.org/abs/https://doi.org/10.1021/nl402455n}
  {https://doi.org/10.1021/nl402455n} \BibitemShut {NoStop}%
\bibitem [{\citenamefont {Peierls}(1933)}]{Peierls-ZPhys-1933}%
  \BibitemOpen
  \bibfield  {author} {\bibinfo {author} {\bibfnamefont {R.}~\bibnamefont
  {Peierls}},\ }\href@noop {} {\bibfield  {journal} {\bibinfo  {journal} {Z.
  Phys.}\ }\textbf {\bibinfo {volume} {80}},\ \bibinfo {pages} {763} (\bibinfo
  {year} {1933})}\BibitemShut {NoStop}%
\end{thebibliography}

\end{document}